\documentclass[ 
twocolumn,
showpacs,preprintnumbers,
bibnotes,
 amsmath,amssymb,
 aps,
 pra,
superscriptaddress,
longbibliography,
]{revtex4-1}

\usepackage[dvipsnames]{xcolor}
\usepackage{textcomp}
\usepackage{graphicx}
\usepackage{subfigure}
\definecolor{pacificb}{HTML}{1CA9C9}
\usepackage[normalem]{ulem}
\newcommand{\iu}{{i\mkern1mu}}
\usepackage{physics}
\makeatletter
\newcommand\incircbin
{%
  \mathpalette\@incircbin
}
\newcommand\@incircbin[2]
{%
  \mathbin%
  {%
    \ooalign{\hidewidth$#1#2$\hidewidth\crcr$#1\ovoid$}%
  }%
}

\makeatother

\newcommand{\bothship}[2]{\langle#1\text{\textbar}#2\rangle}
\newcommand{\tripleship}[3]{\langle#1\text{\textbar}#2\text{\textbar}#3\rangle}

\begin{document}

\title{A trion in a magnetic field revisited}%

\author{I. A. Aleksandrov}
\affiliation{Department of Physics, Saint Petersburg State University, Universitetskaya Naberezhnaya 7/9, Saint Petersburg 199034, Russia}
\affiliation{Ioffe Institute, Politekhnicheskaya street 26, Saint Petersburg 194021, Russia}
\email{i.aleksandrov@spbu.ru}

\author{A. Kudlis}
\affiliation{Abrikosov Center for Theoretical Physics, MIPT, Dolgoprudnyi, Moscow Region 141701, Russia}
\email{kudlis.a@mipt.ru}

\author{I. A. Shelykh}
\affiliation{Science Institute, University of Iceland, Dunhagi-3, IS-107 Reykjavik, Iceland}
\affiliation{Abrikosov Center for Theoretical Physics, MIPT, Dolgoprudnyi, Moscow Region 141701, Russia}
\email{shelykh@hi.is}

\date{\today}

\begin{abstract}
We revisit the problem of a two dimensional trion in an external magnetic field. We demonstrate that the approximations used previously for finding the energy spectrum of this system break down in the experimentally accessible range of magnetic fields. It is shown that the neglect of the Coulomb-induced mixing of different Landau levels corresponding to non-interacting particles leads to a strong underestimation of the trion binding energies even at extremely high magnetic fields (hundreds of Tesla).  Moreover, proper account of the Coulomb effects for certain values of the parameters can lead to the appearance of additional discrete trion states which were overlooked previously. Finally, we provide a database of the matrix elements necessary for calculation of the magnetotrion spectra for a wide class of materials.
\end{abstract}

\maketitle
\section{Introduction}\label{sec:introduction}
\allowdisplaybreaks

Understanding of magneto-optical spectra of various quasi-two-dimensional (2D) electron-hole ($e$--$h$) complexes in external magnetic fields is an important problem in condensed matter theory~\cite{Miura2008,10.1063/5.0042683}. The resonances corresponding to such composite quasiparticles in magneto-optical spectra are routinely observed in semiconductor quantum wells and 2D materials.
These experimental observations give strong motivation for theorists to provide a corresponding quantitative description together with an unambiguous classification of the spectra. The developed theoretical techniques should be sufficiently flexible and allow one to examine a broad variety of materials with different effective masses of carriers and types of the screening in experimentally relevant diapasons of magnetic fields.

The fundamental basis for theoretical description of magneto-optic response was laid by L.~D.~Landau decades ago in his seminal work introducing the concept of Landau levels (LLs) for a single charged particle in an external homogeneous magnetic field~\cite{Landau1930}. The next step was consideration of an exciton, a neutral composite particle consisting of an electron and a hole bound together by the Coulomb attraction~\cite{Akimoto1967,PhysRevB.39.7697,PhysRevB.33.8336,PhysRevB.43.14707,PhysRevLett.58.2598,PhysRevB.95.195311}. In this case, it was demonstrated that center-of-mass coordinates of an exciton can be separated from the internal degrees of freedom~\cite{Gorkov_Dzyaloshinskii_1968,AvronHerbst_1978}. This allowed one to build the theory of a magnetoexciton using a perturbative approach~\cite{Lozovikex_1980}, WKB approximation~\cite{Akimoto1967}, and brute force numerical modelling~\cite{PhysRevB.39.7697}, and provide the classification of the corresponding spectra~\cite{PhysRevB.33.8336}. 

Besides excitons, electron-hole complexes containing more than two particles, such as biexcitons~\cite{He2016,Stevens2018,Torche2021} and trions~\cite{Finkelstein1995,PhysRevB.52.7841,PhysRevLett.89.246805,PhysRevLett.123.097403,PhysRevB.62.16835,PhysRevB.72.165339,PhysRevB.105.L041302,PhysRevLett.129.076801,PhysRevX.10.021024,Li2020} can be formed. The latter case is of particular interest as a trion ($\textup{T}$), being a composite particle consisting of two holes and one electron  ($\textup{X}^{+}$) or two electrons and one hole ($\textup{X}^{-}$), possesses a net electric charge, and its center-of-mass dynamics should be thus strongly affected by an external magnetic field~\cite{PhysRevB.35.4331,PhysRevB.57.1762,PhysRevLett.74.976,PhysRevB.51.10880,PhysRevB.54.R2296,PhysRevB.56.15185,PhysRevB.58.9926,PhysRevB.56.2044,PhysRevB.59.2927,RIVA2002543,PhysRevB.63.115302,PhysRevB.98.205303,Filikhin_2018,10.1063/5.0096834,10.1063/5.0057493,PhysRevB.61.2888,KAUR2019347}. Although for the case of a magnetotrion the center-of-mass coordinates cannot be straightforwardly separated as in the case of an exciton, one can take advantage of an another exact symmetry, namely magnetic translations, to reduce the number of the degrees of freedom~\cite{PhysRevA.43.287}. An additional quantum number characterizing trion states can be thus introduced, together with the total orbital angular momentum $M_z$ along the magnetic-field direction, and the spins of electrons $S_e$ and holes $S_h$~\cite{PhysRevLett.84.4429}.

A number of different techniques were employed to analyze magnetotrion spectra quantitatively, such as variational methods with Slater-type orbitals as trial functions~\cite{10.1063/5.0057493,10.1063/5.0096834} and a stochastic variational approach~\cite{RIVA2002543,PhysRevB.63.115302}. These methods, however, possess an important drawback as they do not present a clear classification of the energy levels. Another possible approach is based on the perturbative treatment where the mixing of the LLs due to the the Coulomb interaction is supposed to be small~\cite{PhysRevLett.84.4429,DZYUBENKO2000683,PhysRevB.65.035318,PhysRevB.98.205303}. However, as we will show below, this approach neglects an entire class of the important matrix elements, and not only gives wrong values of trion energies, but even fails to reproduce the correct number of the discrete states in the experimentally relevant range of magnetic fields. Revisiting the problem is thus a relevant task.

When considering a trion in an external magnetic field, one should analyze the regimes of strong and weak fields  separately. Indeed, on the one hand, in the case of a weak magnetic field, the simplest natural approximation would be to consider a trion as a single particle with the net charge $Q_\textup{T} = e$ and a mass equal to the total mass of the particles forming it, $m_\textup{T} = 2m_e+m_h$. In this limit, the Coulomb interaction binding the particles together dominates over the magnetic field effects, so that
\begin{equation}
    E_{\textup{T}}\gg\hbar\omega_\textup{T} = \frac{\hbar |e| B}{m_\textup{T}c},
    \label{OmegaT}
\end{equation}
where $E_\textup{T}$ is the binding energy of a trion and $\omega_\textup{T} = |e|B/(m_\textup{T}c)$ is the trion cyclotron frequency.

On the other hand, in the case of strong magnetic fields, one can consider three particles interacting with a magnetic field as almost independent ones. It was argued that the Coulomb interaction in this case can be treated as a perturbation leading to a weak mixing of the LLs corresponding to the individual particles. This approach is expected to be valid if~\cite{PhysRevLett.84.4429,DZYUBENKO2000683,PhysRevB.65.035318}
\begin{equation}
\hbar \omega_e, \, \hbar \omega_h, \, |\hbar \omega_e -\hbar \omega_h| \gg  \frac{e^2}{\varepsilon \lambda}=\frac{e^2}{\varepsilon}\sqrt{\frac{|e|B}{\hbar c}},\label{criterion2}
\end{equation}
where $\omega_e$ and $\omega_h$ are the cyclotron frequencies of an electron and a hole, respectively, $\varepsilon$ is a permittivity of a medium, and $\lambda = \sqrt{\hbar c/(|e|B)}$, with  is a magnetic length. This corresponds to fields
\begin{equation}
B\gg \frac{|e|^3 m_{e,h}^2 c }{\varepsilon^2 \hbar^2}.\label{Bmin}
\end{equation} 
Here we employ the Gaussian-cgs units (the SI expressions can be obtained by substituting $\varepsilon \to 4 \pi \varepsilon \varepsilon_0$ and omitting $c$). The condition~\eqref{criterion2} seems to have a clear physical meaning, being equivalent to the requirement that the characteristic mixing energy between the LLs were much smaller than the distance between them.

However, one can note that there are certain problems with this criterion. First,  Eq.~\eqref{Bmin} imposes a very strong restriction on the values of the magnetic field strength. For instance, in the case of GaAs, it corresponds to $B \sim 1000~\textup{T}$ or stronger, which is out of the experimentally accessible range. 

Second, even if the condition~\eqref{Bmin} is satisfied, one can argue that the mixing of the LLs should be taken into account anyway. Indeed, from the viewpoint of the perturbation theory, the corresponding second-order energy correction is proportional to the square of the Coulomb matrix element (which scales as $\sqrt{B}$), divided by the distance between the LLs (which is proportional to $B$). This combination is thus independent on the magnetic field, which means that the increase of the latter does not reduce the role of the mixing terms. 

These two considerations motivated us to go beyond the approximation used in Refs.~\cite{PhysRevLett.84.4429,DZYUBENKO2000683,PhysRevB.65.035318} and treat the Coulomb interaction terms in the magnetotrion problem exactly. In this paper, we present the corresponding development of the operator method introduced in Refs.~\cite{PhysRevLett.84.4429,DZYUBENKO2000683,PhysRevB.65.035318} generalizing it for the case when mixing of the LLs is accounted for, and provide a reader with all necessary formulae for all relevant matrix elements. We show that the approach used previously considerably underestimates the trion binding energy and, moreover, fails to reproduce completely the discrete part of the spectrum. 

The paper has the following structure. In Sec.~\ref{sec:theory}, we briefly outline our theoretical approach (detailed derivations are given in the Appendices). In Sec.~\ref{sec:results}, we provide a quantitative analysis of trion energy spectra in the cases of GaAs and CdTe quantum wells (QWs) for various values of the magnetic field strength. Conclusions are summarized in Sec.~\ref{sec:conclusion}.

Throughout the text, we denote the electron (hole) charge and mass by $e < 0$ and $m_e$ ($-e$ and $m_h)$.

\section{Theoretical description} \label{sec:theory}

In this section, we discuss the main theoretical aspects of the calculations of trion energy spectra. We briefly describe the symmetries of the problem, the basis used, and present the final expressions for the matrix elements of the full trion Hamiltonian. In the main text we present only the most important steps of the derivation, whereas detailed description of the calculation procedures is given in the Appendices. We tried to make the paper self-consistent, so Appendices contain all relevant expressions for the cases of individual electrons, excitons, trions, and unbound electron-exciton pairs placed in external magnetic fields (see Appendix~A, B, C, and D, respectively). 

The Hamiltonian of a three-particle system involving a pair of two dimensional (2D) electrons with coordinates $\vb{r}_1$ and $\vb{r}_2$ and a hole with coordinates $\vb{r}_h$ reads
\begin{equation}
\hat{H}^{(\textup{T})} = \hat{H}_0 + \hat{V}_{eh} + \hat{V}_{ee},
\label{eq:Ham_b}
\end{equation}
where
\begin{align}
\hat{H}_0 &= \sum_{j=1,2} \frac{\hat{\boldsymbol{\pi}}_j^2}{2m_e} + \frac{\hat{\boldsymbol{\pi}}_h^2}{2m_h}, \\
\hat{V}_{eh} &=  - \sum_{j=1,2} \frac{e^2}{ \lambda} V(|\vb{r}_j - \vb{r}_h|), \\
\hat{V}_{ee} &= \frac{e^2}{ \lambda} V(|\vb{r}_1 - \vb{r}_2|).
\end{align}
Here the momenta operators are obtained by the canonical substitution $\hat{\boldsymbol{\pi}}_j = -i \hbar \boldsymbol{\nabla}_j - (e/c) \boldsymbol{A} (\boldsymbol{r}_j)$, $\hat{\boldsymbol{\pi}}_h = -i \hbar \boldsymbol{\nabla}_h + (e/c) \vb{A} (\vb{r}_h)$, $\lambda = \sqrt{\hbar c/(|e|B)}$, and $V(r)$ is a dimensionless function describing the interaction potential. In the current paper, we focus on the case of conventional semiconductor QWs and thus employ the standard expression for a Coulomb potential,
\begin{equation}\label{eqn:st_Coul_pot}
V(r)=\frac{\lambda}{\varepsilon r}.   
\end{equation}

We did not wright explicitly the terms, correspopnding to the direct action of the magnetic field on spins of individual electrons and holes, as their account is trivial. 

In contrast to the case of a single charged particle (an electron or a hole), where the spectrum is given by the discrete harmonic oscillator series (LLs), the spectrum of a multi-particle complex is substantially modified by the Coulomb interaction, and the problem cannot be solved analytically. Therefore, we first construct the basis consisting of the eigenfunctions of the free particle Hamiltonian $\hat{H}_0$ and then diagonalize the matrix of the total Hamiltonian $\hat{H}^{(\textup{T})}$ numerically. When choosing the basis, we fully take into account the complete symmetries of the system, which allows us to substantially reduce the calculation costs. 

In the case of the three interacting particles, the operator of the {\it magnetic translations} defined as~\cite{Lozovikex_1980,PhysRevLett.84.4429, DZYUBENKO2000683, PhysRevB.65.035318}
\begin{equation}
\hat{\vb{K}} = \sum_{j=1,2} \hat{\boldsymbol{\pi}}_j + \hat{\boldsymbol{\pi}}_h + \frac{e}{c} \sum_{j=1,2} \vb{B} \times \vb{r}_j - \frac{e}{c} \, \vb{B} \times \vb{r}_h
\label{eq:K_eeh_b}
\end{equation}
commutes with the Hamiltonian $\hat{H}^{(\textup{T})}$.  Note that this commutator is proportional to the net electric charge of the system, so the components of the corresponding operator for an exciton do commute (see Appendix~\ref{sec:app_exciton}). However, the $x$ and $y$ components of $\hat{\vb{K}}$ do not commute with each other, $[\hat{K}_x, \, \hat{K}_y ] = i|e|\hbar B /c$, which means that in the case of a trion, when constructing the basis, one can take advantage of only one integral of motion (for instance, either $\hat{K}_x$ or $\hat{K}_y$). Instead of choosing a specific component of the vector $\hat{\vb{K}}$ (either $\hat{K}_x$ or $\hat{K}_y$), one can use the operator $\hat{\vb{K}}^2$~\cite{DZYUBENKO2000683}, which commutes with the Hamiltonian as well. 

Let us perform the following coordinate transformation:
\begin{align}
\vb{R} &= \frac{1}{\sqrt{2}} \, \big ( \vb{r}_1 + \vb{r}_2 \big ), \quad \vb{r} = \frac{1}{\sqrt{2}} \, \big ( \vb{r}_1 - \vb{r}_2 \big ).
\end{align}
In the most straightforward approach the basis functions corresponding to the case of non-interacting particles can be chosen as:
\begin{multline}
\varphi^{(\textup{T})}_{n_1 m_1 n_2 m_2 n_h m_h} (\vb{r}, \vb{R}, \vb{r}_h) \\ 
= \varphi^{(e)}_{n_1 m_1} (\vb{r}) \varphi^{(e)}_{n_2 m_2} (\vb{R}) \varphi^{(h)}_{n_h m_h} (\vb{r}_h),
\label{eq:eeh_naive_b}
\end{multline}
where $\varphi^{(h)}_{n m} (\vb{r}) = [ \varphi^{(e)}_{n m} (\vb{r}) ]^*$ correspond to single particle wavefunctions in a magnetic field (see Appendix~\ref{sec:app_one_electron}).  The quantum number $n = 0, 1, 2, \dots$ is the LL number, and $m = 0, 1, 2, \dots$ defines the angular momentum projection. However, this basis set completely disregards the translational symmetry of the system as the corresponding functions are characterized only by a well-defined projection of the total angular momentum $M_z =(n_1-m_1)+(n_2-m_2)-(n_h-m_h)$ and total electron spin $S_e$ [$S_e = 0$ (spin-singlet state) for even $n_1 - m_1$ and $S_e = 1$ (spin-triplet state) for odd $n_1 - m_1$; see Appendix~\ref{sec:app_trinoi_prob}]. 

To account for the translational symmetry, one can perform a unitary transformation making the basis functions~\eqref{eq:eeh_naive_b} also eigenfunctions of $\hat{\vb{K}}^2$~\cite{DZYUBENKO2000683} (see Appendix~\ref{sec:app_trinoi_prob} for details), and get the following set of the basis functions  
\begin{widetext}
\begin{align}
    \psi_{n_1 n_2 n_h m l}(\vb{r},\vb{R},\vb{r}_h)&=\frac{1}{\sqrt{N_{n_1n_2n_hml}}}\left[\dfrac{1}{\sqrt{2}}\left(\xi-\pdv{}{\xi^*}\right)\right]^{n_1}\left[\dfrac{1}{\sqrt{2}}\left(\xi_R-\pdv{}{\xi^*_R}\right)\right]^{n_2}\left[\dfrac{1}{\sqrt{2}}\left(\xi^*_h-\pdv{}{\xi_h}\right)\right]^{n_h}\psi_{ml}(\vb{r},\vb{R},\vb{r}_h), \label{eq:psi_basis}
\end{align}
\end{widetext}
where
\begin{align}
  \psi_{ml}(\vb{r},\vb{R},\vb{r}_h)=\left(\xi^*\right)^m\left(\xi_h\right)^l\tilde{\Phi}^{(\textup{T})}_0\!(\vb{r},\!\vb{R},\vb{r}_h)
\end{align}
and  
\begin{align}
\tilde{\Phi}^{(\textup{T})}_0\!(\vb{r},\!\vb{R},\vb{r}_h)\!=\!\dfrac{\exp{-\xi\xi^*\!-\!\xi_R\xi^*_R\!-\!\xi_h\xi^*_h\!+\!\sqrt{2}\xi_h\xi^*_R}}{\sqrt{2}\left(2\pi\right)^{3/2}\lambda^3}.
\end{align}
is a normalized wave function of the ground state. The normalization factors $N_{n_1n_2n_hml}$ are calculated in Appendix~\ref{sec:app_trinoi_prob}. The dimensionless coordinates read: $\xi=z/(2\lambda)$, $\xi_R=z_R/(2\lambda)$, and $\xi_h=z_h/(2\lambda)$, where $z=x+\iu y$.  The angular momentum projection of the state~\eqref{eq:psi_basis} reads
\begin{equation}
    M_z=n_1-m+n_2-(n_h-l),
\end{equation}
while the spin quantum number is $S_e = 0$ ($S_e = 1$) if $n_1 - m$ is even (odd). Note that the energies are degenerate with respect to the quantum number $k$ associated with the operator $\vb{K}^2$, so we have set $k=0$.

The matrix elements of the full Hamiltonian~\eqref{eq:Ham_b} are defined by the following expression:
\begin{align}
H^{(\textup{T})}_{n_1 n_2 n_h m l; n_1' n_2' n_h' m' l'} &= E^{(\textup{T},0)}_{n_1+n_2, n_h} \delta_{n_1 n_1'} \delta_{n_2 n_2'} \delta_{n_h n_h'} \delta_{m m'} \delta_{l l'} \nonumber \\
{} & + V^{(eh)}_{n_1 n_2 n_h m l; n_1' n_2' n_h' m' l'} \nonumber \\
{} &+ V^{(ee)}_{n_1 n_2 n_h m l; n_1' n_2' n_h' m' l'} ,
\label{eq:HT_b}
\end{align}
where 
\begin{equation} \label{FreeLL}
E^{(\textup{T},0)}_{n_e,n_h} = \hbar \omega_e \bigg( 1 + n_e \bigg) + \hbar \omega_h \bigg( \frac{1}{2} + n_h \bigg)
\end{equation}
corresponds to the LLs of non-interacting particles, $n_e=n_1+n_2$ and $\omega_{e,h} = |e|B/(m_{e,h}c)$. The Coulomb interaction matrix elements $V^{(eh)}_{n_1 n_h m l; n_1' n_2' n_h' m' l'}$ and $V^{(ee)}_{n_1 n_2 n_h m l; n_1' n_2' n_h' m' l'}$ are bulky and given in Appendic C ( Eqs.~\eqref{eqn:app_matr_el_eh} and \eqref{eqn:app_matr_el_ee}, respectively).

In Ref.~\cite{DZYUBENKO2000683} the Coulomb mixing was taken into account only within the subspaces corresponding to $n_e=n_e´$ and $n_h=n_h´$, i.e. it was thus assumed, that $n_e,n_h$  remain good quantum numbers of the full interaction problem. It was argued that this approximation should be valid in the case of very strong magnetic fields, where the magnetic energy dominates over the Coulomb energy. However, as was discussed in the Introduction [see Eq.~\eqref{Bmin} and below], this approach can hardly be justified in any physically relevant situation and exact treatment of the Coulomb mixing is necessary.

The resulting spectrum contains discrete levels and continuous regions, corresponding to the states of unbound electron-exciton pairs. To identify the onset of the energy continuum, we additionally perform the independent calculations for the case of an uncoupled electron and an exciton (see Appendix~\ref{sec:X_e}). 
\begin{figure}
    \centering
    \includegraphics[width = 1.0\linewidth]{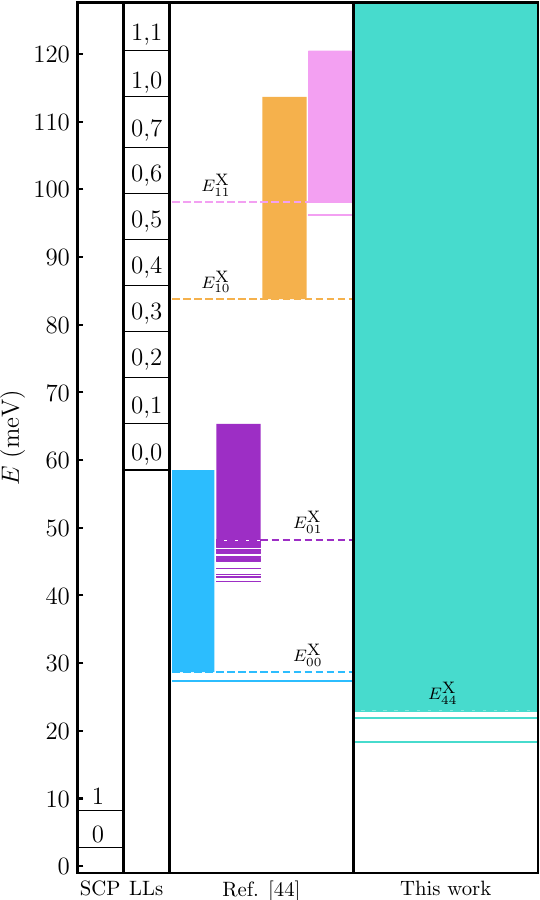}
    \caption{Lower part of the energy spectrum for a trion in a GaAs QW for $B=30~\textup{T}$. The first vertical panel corresponds to the-single-composite  particle (SCP) approximation, where the energies are calculated by means of Eq.~\eqref{eqn:SP_ener}. The second vertical panel displays the free LLs obtained via Eq.~\eqref{FreeLL}. Near each horizontal line, we indicate the values of the pair $(n_e=n_1+n_2, n_h)$. In the third panel, we present the results obtained with use of the technique proposed in Ref.~\cite{PhysRevLett.84.4429}, where Coulomb mixing of different LLs was neglected.  The fourth panel displays the spectrum obtained in this work by diagonalization of the matrix of the full Hamiltonian, \eqref{eq:HT_b}, accounting for the LLs mixing. In the third and fourth panels thin horizontal lines correspond to the discrete energy state of magnetotrions, and color bars display energy continuums corresponding to unbound electron-hole pairs. As one can see, even for very strong value of the magnetic field  the account of the Coulomb mixing is crucial. It affects substantially both the boundary of the energy continuum and magnetotrion binding energy, defined as the distance between the discrete trion state and the boundary of energy continuum. Moreover, it also leads to the appearance of a second discrete level.
    To make the figure less cumbersome, we have not plotted all regions of the energy continuums in the third panel, which should start down from every LL of non-interacting particles presented in the second panel.}
    \label{fig:pic1_gaAs_30}
\end{figure}

\section{Results and discussion} \label{sec:results}

In this section, we examine the trion spectra for GaAs and CdTe QWs for different values of the external magnetic field strengths. For these conventional semiconductors, the interaction among charge carriers is well described by the standard Coulomb potential~\eqref{eqn:st_Coul_pot}.

For the case of GaAs, we use the following parameters: effective masses of carriers $m_e = 0.063 m_0$, $m_h = 0.51 m_0$ ($m_0$ is the free electron mass), static dielectric permittivity $\varepsilon = 12.9$. 

Our results for $B = 30~\textup{T}$ are shown in Fig.~\ref{fig:pic1_gaAs_30}. The corresponding spectrum is presented in the fourth vertical panel from the left. Constructing the matrix of the Hamiltonian, we accounted for the LLs with $n_{h}$ and $n_{e}=n_1+n_2$ spanning from $0$ to $4$ (see the discussion of the convergence of the procedure in Appendix G). We compared our results with those obtained in Refs.~\cite{PhysRevLett.84.4429, DZYUBENKO2000683, PhysRevB.65.035318}, where the Coulomb mixing of the different LLs was neglected, see third vertical panel from the left. The thin horizontal lines correspond to the discrete energy state of magnetotrions, and the color bars display the energy continua corresponding to unbound electron-hole pairs. For comparison we also illustated the simple cases of a trion
as a single composite particle (first panel from the
right)with the spectrum given by:
\begin{align}\label{eqn:SP_ener}
    E^{\textup{SCP}}_n= \hbar \omega_\textup{T} \left(n+\dfrac{1}{2}\right),
\end{align}

where the trion cyclotron frequency $\omega_T$ is defined in Eq.~\eqref{OmegaT}, and LLs corresponding to three non-interacting particles with spectrum given by Eq.~\eqref{FreeLL} (second panel from the left). Both of these approximations yield obviously only discrete energy levels.

Comparing third and fourth panels of Fig.~\ref{fig:pic1_gaAs_30} one can note that even for the value of a magnetic field equal to several tens of Tesla, the account of the Coulomb mixing is crucial. It substantially affects the boundary of the energy continuum, which shifts down by $5.688~\textup{meV}$). The magnetotrion binding energy, defined as the distance between the discrete trion state and the boundary of the energy continuum, is increased from $1.298~\textup{meV}$ ($M_z=-1$ and $S_e=1$) to $4.680~\textup{meV}$ ($M_z=0$ and $S_e=0$). For details, see Appendices~\ref{sec:app_cont_onset} and~\ref{app:sec_fill_ham}.

\begin{figure}[t!]
    \centering
    \includegraphics[width = 1.0\linewidth]{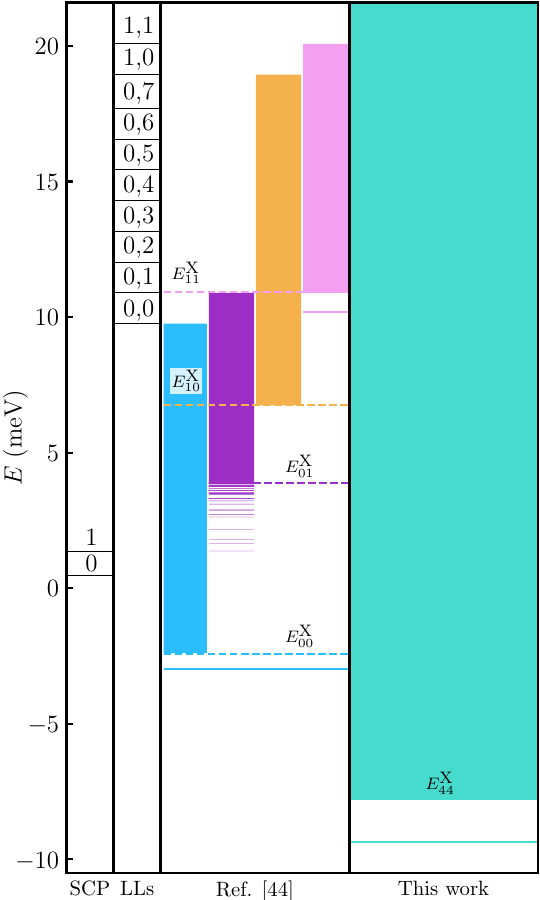}
        \caption{Lower part of the energy spectrum for a trion in a GaAs QW for $B=5~\textup{T}$. The first vertical panel corresponds to the-single-composite  particle (SCP) approximation, where the energies are calculated by means of Eq.~\eqref{eqn:SP_ener}. The second vertical panel displays the free LLs obtained via Eq.~\eqref{FreeLL}. Near each horizontal line, we indicate the values of the pair $(n_e=n_1+n_2, n_h)$. In the third panel, we present the results obtained with use of the technique proposed in Ref.~\cite{PhysRevLett.84.4429}, where Coulomb mixing of different LLs was neglected.  The fourth  panel displays the spectrum obtained in this work by diagonalizing the matrix of the full Hamiltonian, \eqref{eq:HT_b}, accounting for the LLs mixing. In the third and fourth panels thin horizontal lines correspond to the discrete energy state of magnetotrions, and color bars display energy continuums corresponding to unbound electron-hole pairs. One clearly sees the substantial shift of both the energy continuum and the trion binding energy. }
    \label{fig:pic3_gaAs_5}
\end{figure}

Moreover, exact treatment of the Coulomb mixing of LLs modifies the spectrum quantitatively,
leading to the appearence of an additonal discrete trion 
state below the energetic continuum with binding energy of $1.043~\textup{meV}$. This state corresponds to $M_z=-1$ and $S_e=1$.

We also considered the the case of $B=5~\textup{T}$, which is illustrated in Fig.~\ref{fig:pic3_gaAs_5}. Here the energy continuum moves down by $5.329~\textup{meV}$, and the trion binding energy increases from $0.530~\textup{meV}$ to $1.604~\textup{meV}$. Interestingly, for this value of the magnetic field, an additional trion level does not appear.

\begin{figure}[t!]
    \centering
    \includegraphics[width = 1.0\linewidth]{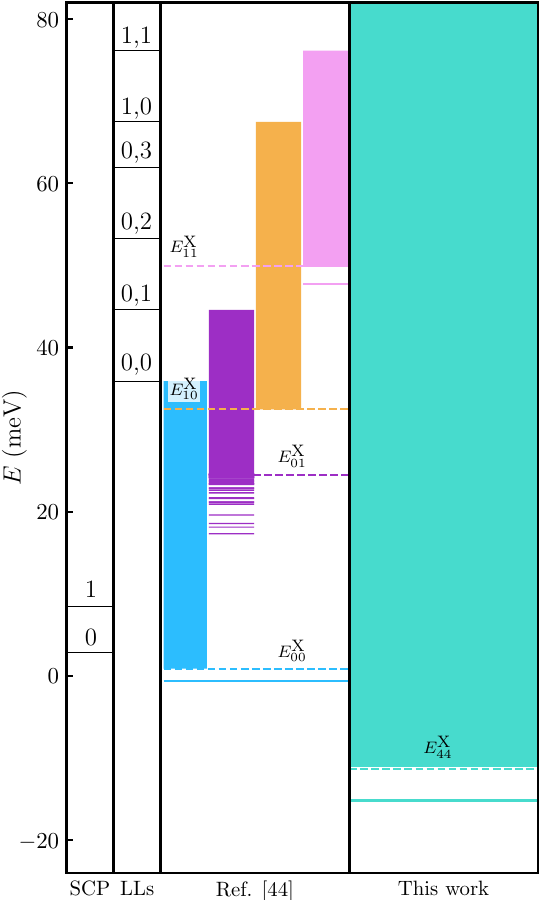}
    \caption{Lower part of the energy spectrum for a trion in a CdTe QW for $B=30~\textup{T}$. The first vertical panel corresponds to the-single-composite  particle (SCP) approximation, where the energies are calculated by means of Eq.~\eqref{eqn:SP_ener}. The second vertical panel displays the free LLs obtained via Eq.~\eqref{FreeLL}. Near each horizontal line, we indicate the values of the pair $(n_e=n_1+n_2, n_h)$. In the third panel, we present the results obtained by means of the technique proposed in Ref.~\cite{PhysRevLett.84.4429}, where Coulomb mixing of different LLs was neglected.  The fourth  displays the spectrum obtained in this work by diagonalizing the matrix of the full Hamiltonian, \eqref{eq:HT_b}, accounting for the LLs mixing. In the third and fourth panels thin horizontal lines correspond to the discrete energy state of magnetotrions, and color bars display energy continuums corresponding to unbound electron-hole pairs. One clearly sees the substantial shift of both the energy continuum and the trion binding energy.}
    \label{fig:pic4_CdTe_30}
\end{figure}

For comparison, we also present the results for another widely used semiconductor material, CdTe. It has effective masses $m_e = 0.11 m_0$, $m_h = 0.40 m_0$ and relative permittivity $\varepsilon = 11.0$. We show the spectrum for $B = 30~\textup{T}$ only (see Fig.~\ref{fig:pic4_CdTe_30}). As in the case of GaAs, our findings demonstrate that  proper account for the Coulomb mixing of the LLs is important and leads to the shift of the energy continuum by $11.796~\textup{meV}$ and of the trion binding energy by $2.735~\textup{meV}$.  

\begin{figure*}[t!]
\centering
\includegraphics[width = 0.8\linewidth]{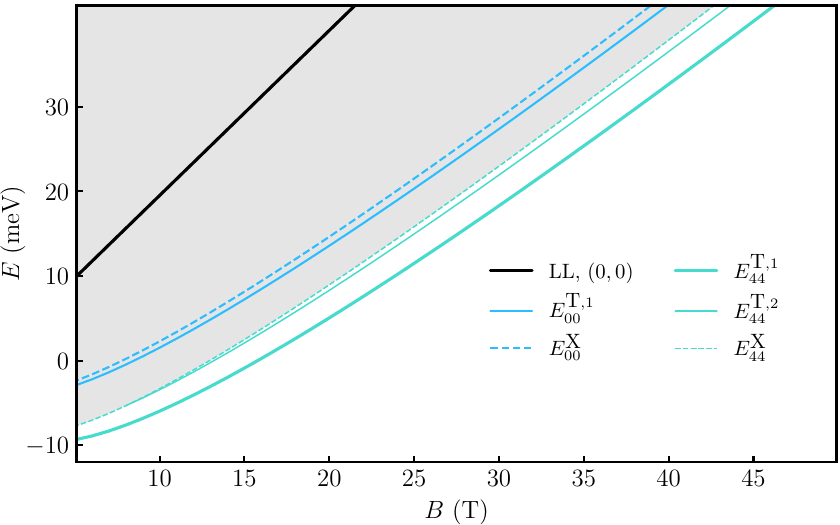}
\caption{The energy spectrum of GaAs based system as function of the magnetic field $B$. Solid green lines correspond to the two discrete trion states obtained in this work, the dashed green line corresponds to the lower boundary of the energy continuum, solid blue line corresponds to the trion bound state reported in Ref.~\cite{DZYUBENKO2000683}, dashed blue line indicates the corresponding energetic continuum boundary. We also plot the position of the lowest LL corresponding to non-interacting particles (solid black line). 
The increase of magnetic field does not lead to the merging of green and blue lines, which means that the role of the Coulomb mixing between different LLs does not decrease with magnetic field. We accounted for LLs with $n_{h}$ and $n_{e}=n_1+n_2$ spanning from $0$ to $4$. In Appendix~\ref{app:sec_extrapol}, we also present the similar dependence extracted on the basis of extrapolating procedure described there (see Fig.~\ref{fig:pic6_GaAs_levels_vs_B}).}
\label{fig:pic6_GaAs_levels_vs_B_b}
\end{figure*}

In Fig.~\ref{fig:pic6_GaAs_levels_vs_B_b} we show the  energy spectrum of the system as a function of the magnetic field $B$. We display the positions of the two discrete trion states, which were identified in our calculations (thin and thick solid green lines), the lower boundary of the energy continuum (dashed green line) and compare them to the results, presented in Ref.~\cite{DZYUBENKO2000683} shown by the solid blue line (bound trion) and dashed line (continuum boundary). The position of the lowest LL corresponding to non-interacting particles is represented by the solid black line.

As one can clearly see, the increase of the magnetic field does not lead to the merging of green and blue lines, which means that the role of the Coulomb mixing between the different LLs does not decrease with increasing~$B$. This is in agreement with a qualitative argument that although the increase of the magnetic field leads to LL separation, it also enhances the Coulomb matrix elements due to the reduction of cyclotron radii ($\lambda \sim 1/\sqrt{B}$), and these two effects compensate each other (see the discussion after Eq.~\eqref{Bmin} in the Introduction). 

We point out once more that in our calculations the spin structure of the wave functions was not presented explicitly. The electron spin state $S_e$ only defines the parity of the coordinate part of the wave function. The account of the interaction of spin with magnetic field is trivial, and will lead to the splitting of a state with $S_e$ into $(2S_h+1)(2S_e + 1)=2(2S_e + 1)$ sublevels. 

\section{Conclusion}\label{sec:conclusion}

In this study, we revisited the problem of a 2D trion in an external magnetic field. We demonstrated, that the  mixing between different Landau levels induced by Coulomb interaction plays crucial role in the determination of the energy spectrum of the problem and can not be neglected even in the limit of very strong magnetic fields. This is because magnetic field increases not only the distance between the Landau levels, but Coulomb matrix elements as well via decrease of the magnetic length. We have shown that proper account of the Coulomb mixing not only leads to the substantial increase of magnetotrion binding energy, but can also lead to qualitative reshaping of the spectrum, resulting in the appearance of additional discrete energy states. 

\section*{Acknowledgments}

I.A.S. acknowledge financial support from Icelandic Research Fund (Rannis, the project ``Hybrid
polaritonics''), ``Priority 2030 Academic Leadership Program'' and ``Goszadanie no. 2019-1246''.

\appendix

\section{Single electron in a magnetic field} \label{sec:app_one_electron}

Here and in Appendices~\ref{sec:app_exciton} and \ref{sec:app_trinoi_prob}, we first briefly outline how the energy spectrum can be computed and then provide a detailed theoretical description of the corresponding problems. In Appendix~\ref{sec:X_e}, we give a direct recipe for calculating the spectrum for an unbound electron-exciton pair, since we presented all the necessary theoretical calculations in the previous sections.

\subsection{Energy spectrum}

In the case of a single electron (hole) in a magnetic field, the energy is given by a simple formula describing the spectrum of a harmonic oscillator. The corresponding wave functions $\varphi^{(e)}_{nm} (\vb{r})$ are constructed explicitly and form a basis in which the Hamiltonian is diagonal:
\begin{align}
H^{(e)}_{n';n} = E^{(e)}_{n} \delta_{nn'},
\end{align}
where $n, n' = 0, 1, 2, \dots$ and the eigenvalues correspond to the Landau levels (LLs),
\begin{equation}
E_n^{(e)} = \hbar \omega_e \bigg( \frac{1}{2} + n \bigg).
\label{eq:1e_energy}
\end{equation}
Here $\omega_e = |e|B/(m_e c)$ is the electron cyclotron frequency. The energy is degenerate with respect to the second quantum number $m$. The $z$ projection of the orbital angular momentum is $m_z = n - m$.

\subsection{Details}\label{app:sec_one_electron}

We start with the Hamiltonian, which in the case of a single electron interacting with a constant uniform magnetic field reads
\begin{equation}
\hat{H}^{(e)} = \frac{\hat{\boldsymbol{\pi}}_e^2}{2m_e},
\label{eq:H_1e}
\end{equation}
where $\hat{\boldsymbol{\pi}}_e = -i \hbar \boldsymbol{\nabla} - (e/c) \vb{A} (\vb{r})$ and we employ a symmetric gauge, $\vb{A} (\vb{r}) = (\vb{B} \times \vb{r})/2$ with $\vb{B}$ directed along the $z$ axis. We assume that the particle is confined within the $xy$ plane and its quantum state is described by a wave function depending solely on $x$ and $y$ (or polar coordinates $r$ and $\varphi$).

In the external field specified above, the Hamiltonian~\eqref{eq:H_1e} obviously commutes with the operator of the angular momentum projection $\hat{\ell}_z = -i \hbar \partial_\varphi$, so the eigenfunctions of $H^{(e)}$ will also be characterized by the corresponding quantum number $m_z$, i.e., one can utilize the ansatz $\psi (r, \varphi) = \mathrm{e}^{im_z\varphi} R (r) $. The ordinary differential equation for $R (r)$ can then be solved analytically (see, e.g., Ref.~\cite{LL3}). Here we will use another approach based on the formalism of creation and annihilation operators (see, e.g., Ref.~\cite{DZYUBENKO2000683}).

First, let us introduce the one-particle normalized vacuum state in the coordinate representation:
\begin{equation}
\Phi^{(e)}_0 (\vb{r}) = \langle \vb{r} | 0 \rangle = \frac{1}{\lambda \sqrt{2\pi}} \, \mathrm{e}^{-r^2/(4\lambda^2)},
\label{eq:vacuum_1el}
\end{equation}
where
\begin{equation}
    \lambda = \sqrt{\frac{\hbar c }{|e|B}}
\end{equation}
is the so-called magnetic length, which was also introduced in the main text. It is convenient to use a complex variable $z = r \mathrm{e}^{i\varphi} = x + iy$ and its complex conjugate $z^*$. We will also need the following creation operators in the coordinate representation:
\begin{eqnarray}\label{eqn:cr_an_operators_electron}
\hat{A}_e^\dagger &=& \frac{1}{\sqrt{2}} \bigg ( \frac{z}{2\lambda} - 2 \lambda \, \frac{\partial}{\partial z^*} \bigg ), \label{eq:Ae} \\
\hat{B}_e^\dagger &=& \frac{1}{\sqrt{2}} \bigg ( \frac{z^*}{2\lambda} - 2 \lambda \, \frac{\partial}{\partial z} \bigg ), \label{eq:Be}
\end{eqnarray}
Together with the annihilation operators $\hat{A}_e$ and $\hat{B}_e$, they obey the relations $[\hat{A}_e, \, \hat{A}_e^\dagger] = 1$, $[\hat{B}_e, \, \hat{B}_e^\dagger] = 1$, and $[\hat{A}_e, \, \hat{B}_e] = [\hat{A}_e, \, \hat{B}_e^\dagger] = 0$. These operators allow one to generate the following wave functions:
\begin{equation}
\varphi^{(e)}_{nm} (\vb{r}) = \frac{1}{\sqrt{n!m!}} \, (\hat{A}_e^\dagger)^n (\hat{B}_e^\dagger)^m \Phi^{(e)}_0 (\vb{r}),
\label{eq:phi_1el}
\end{equation}
where $n$, $m = 0$, $1$, $2$, ... One can explicitly verify that these functions form an orthonormal and complete set of the Hamiltonian eigenfunctions and the corresponding energies are given by Eq.~\eqref{eq:1e_energy}. Instead of the non-negative integers $n$ and $m$, one can also label the solutions by the following numbers:
\begin{eqnarray}
\tilde{n} &=& \mathrm{min} \, (n, \, m),  \label{eq:n_tilde} \\
\tilde{m} &=& m - n. \label{eq:m_tilde}
\end{eqnarray}
These allow one to provide the following explicit form of the functions~\eqref{eq:phi_1el}:
\begin{eqnarray}
\varphi^{(e)}_{nm} (\vb{r}) &=& \frac{(-1)^{\tilde{n}}}{\lambda} \sqrt{\frac{2^{|\tilde{m}|} \tilde{n}!}{2 \pi (\tilde{n}+|\tilde{m}|)!}} \bigg ( \frac{r}{2\lambda} \bigg )^{|\tilde{m}|} \nonumber \\
{}& \times & \mathrm{e}^{-i\tilde{m} \varphi} L^{|\tilde{m}|}_{\tilde{n}} \bigg ( \frac{r^2}{2 \lambda^2} \bigg ) \, \mathrm{e}^{-r^2/(4\lambda^2)},
\label{eq:phi_1e_explicit}
\end{eqnarray}
where $L^{|\tilde{m}|}_{\tilde{n}}$ are the Laguerre polynomials. In terms of $n$ and $m$, the explicit form~\eqref{eq:phi_1e_explicit} strongly depends on whether $n>m$ or not, so using $\tilde{n}$ and $\tilde{m}$ is often more convenient. The angular momentum projection reads $m_z = -\tilde{m} = n - m$.

\section{Exciton $(\textup{X})$ in a magnetic field} \label{sec:app_exciton}

In the case of two- and theree-particle systems considered in what follows, the free LLs are modified due to the Coulomb interaction and the Hamiltonian eigenfunctions cannot be found analytically. Next, we will discuss how the energy eigenvalues can be obtained by diagonalizing the corresponding matrix.

\subsection{Spectrum calculation recipe}

The exciton energy spectrum turns out to be striped --- it contains continuous regions separated by empty areas. To evaluate the energies, one has first to separate the center-of-mass motion by introducing a conserved vector $\vb{K}$ (see the next section for details). Using then the basis set~\eqref{eq:eh_wf}, one obtains the matrix elements in the form
\begin{equation}
H^{(\textup{X})}_{\vb{K},n'm'; nm} = E^{(\textup{X},0)}_{nm} \delta_{nn'} \delta_{mm'} + V_{\vb{K},n'm'; nm},
\label{H_matrix_eh}
\end{equation}
where $n$ and $m$ are non-negative integer numbers, $E^{(\textup{X},0)}_{nm}$ is defined in Eq.~\eqref{eq:eh_energy}, while the expression for $V_{\vb{K},n'm'; nm}$ is presented in  Eq.~\eqref{eqn:app_vknm}. The Hamiltonian is block-diagonal in $M_z = n - m$, due to the angular momentum conservation. Having constructed the matrix~\eqref{H_matrix_eh}, we should diagonalize it varying $\vb{K}$ and $M_z$. In fact, the matrix elements depend only on the magnitude $|\vb{K}|$ and $M_z$. Within numerical computations, one has to make sure that for given $|\vb{K}|$ and $M_z$, the quantum numbers $n$ and $m$ change in a sufficiently large range, so that the results converge to a desired accuracy.

\subsection{Details}

As will be seen below, in the case of a two-dimensional electron-hole ($e$--$h$) system, it is possible to separate the center-of-mass coordinates and make use of the one-particle solutions~\eqref{eq:phi_1e_explicit}. The two-particle Hamiltonian has the form
\begin{equation}
\hat{H}^{(\textup{X})} = \frac{\hat{\boldsymbol{\pi}}_e^2}{2m_e} + \frac{\hat{\boldsymbol{\pi}}_h^2}{2m_h} - \frac{e^2}{\lambda} V(|\vb{r}_e - \vb{r}_h|),
\label{eq:H_eh}
\end{equation}
where $\hat{\boldsymbol{\pi}}_e = -i \hbar \boldsymbol{\nabla}_e - (e/c) \vb{A} (\vb{r}_e)$, $\hat{\boldsymbol{\pi}}_h = -i \hbar \boldsymbol{\nabla}_h + (e/c) \vb{A} (\vb{r}_h)$, and $V(r)$ is a dimensionless function governing the interaction potential [for the usual Coulomb interaction without the screening effects, it reads $V(r) = \lambda/(\varepsilon r)$]. Although the interparticle-interaction law has nothing to do with the magnetic length, its typical values does involve $\lambda$ due to the localization of the wave functions in the external magnetic field [for instance, the matrix elements of $V(r) = \lambda/(\varepsilon r)$ computed in the basis of the Landau wave functions are $\lambda$-independent]. 

Let us now take advantage of the translational symmetry of the problem and separate the center-of-mass motion. As was shown in Ref.~\cite{Gorkov_Dzyaloshinskii_1968}, the Hamiltonian~\eqref{eq:H_eh} commutes with the following vector operator:
\begin{eqnarray}
\hat{\vb{K}} &=& \hat{\boldsymbol{\pi}}_e + \hat{\boldsymbol{\pi}}_h + \frac{e}{c} \, \vb{B} \times (\vb{r}_e - \vb{r}_h) \nonumber \\
{} &=& -i \hbar \boldsymbol{\nabla}_e -i \hbar \boldsymbol{\nabla}_h + \frac{e}{c} \, \big [ \vb{A} (\vb{r}_e) - \vb{A} (\vb{r}_h)\big ].
\label{eq:K_eh}
\end{eqnarray}
Since $\hat{K}_x$ and $\hat{K}_y$ also commute with each other, the Hamiltonian eigenfunctions can be characterized by the corresponding vectors $\vb{K}$ consisting of two quantum numbers $K_x$ and $K_y$. Note that in the absence of the external field, $\vb{K}$ is the total momentum of the particles. Let us introduce the center-of-mass position $\vb{R}$ and relative position $\vb{r}$:
\begin{eqnarray}
\vb{R} &=& \frac{m_e \vb{r}_e + m_h \vb{r}_h}{m_e + m_h}, \\
\vb{r} &=& \vb{r}_e - \vb{r}_h.
\end{eqnarray}
The operator~\eqref{eq:K_eh} takes the form
\begin{equation}
\hat{\vb{K}} = -i \hbar \boldsymbol{\nabla}_R + \frac{e}{2c} \, \vb{B} \times \vb{r}.
\end{equation}
The Hamiltonian~\eqref{eq:H_eh} now reads
\begin{eqnarray}
\hat{H}^{(\textup{X})} &=& -\frac{\hbar^2}{2m_{eh}} \, \Delta_R -\frac{\hbar^2}{2\mu} \, \Delta_r \nonumber \\
{}&+& \frac{ie\hbar}{2c} \Bigg \{ \frac{1}{m_{eh}} \big (\vb{B} \times \vb{r} \big ) \boldsymbol{\nabla}_R + \frac{1}{\mu} \Big [ \vb{B} \times \big ( \vb{R} + \gamma \vb{r} \big ) \Big ] \boldsymbol{\nabla}_r \Bigg \} \nonumber \\
{}&+& \frac{e^2 B^2}{8 \mu c^2} \bigg ( R^2 + \beta r^2 + 2 \gamma \vb{r} \vb {R} \bigg ) - \frac{e^2}{ \lambda} V(r),
\end{eqnarray}
where $m_{eh} = m_e + m_h$, $\mu = m_e m_h /m_{eh}$, $\gamma = (m_h - m_e)/m_{eh}$, and $\beta = (m_e^3 + m_h^3)/m_{eh}^3$. The wave function can then be represented as~\cite{Gorkov_Dzyaloshinskii_1968}
\begin{multline}
\varphi^{(\textup{X})}_{\vb{K}} (\vb{r}, \vb{R}) = \mathrm{exp} \bigg [ \frac{i}{\hbar} \bigg ( \vb{K} - \frac{e}{2c} \, \vb{B} \times \vb{r} \bigg ) \vb{R} \bigg ]  \\
{} \times  \mathrm{exp} \bigg ( \frac{i\gamma}{2\hbar} \, \vb{K} \vb{r} \bigg ) \varphi^{(\textup{X})} (\vb{r} - \vb{r}_0).
\end{multline}
It is convenient to choose $\vb{r}_0 = -c(\vb{B} \times \vb{K})/(eB^2)$ in order to obtain the following equation for $\varphi^{(\textup{X})} (\vb{r})$:
\begin{multline}
 \bigg [ - \frac{\hbar^2}{2\mu} \, \Delta + \frac{ie\hbar \gamma}{2\mu c} \, (\vb{B} \times \vb{r}) \boldsymbol{\nabla} + \frac{e^2B^2}{8\mu c^2} \, r^2 \\
 {} - \frac{e^2}{ \lambda} V(|\vb{r} + \vb{r}_0|) - E \bigg ] \varphi^{(\textup{X})} (\vb{r}) = 0.
\label{eq:eh_eq_r}
\end{multline}
Note that the wave function $\varphi^{(\textup{X})} (\vb{r})$ does not involve $\vb{K}$.

First, we will neglect the Coulomb interaction $V$ and find the corresponding two-particle wave functions and energy eigenvalues (free LLs). The total Hamiltonian $\hat{H}^{(\textup{X})}$ will be then considered in this basis, and the corresponding matrix elements are given by Eq.~\eqref{H_matrix_eh}. The solutions of Eq.~\eqref{eq:eh_eq_r} without the $V$ term are exactly those given in Eq.~\eqref{eq:phi_1e_explicit}, so the normalized two-particle wave functions within the zeroth order with respect to the Coulomb interaction have the form
\begin{multline}
\varphi^{(\textup{X})}_{\vb{K},nm} (\vb{r}, \vb{R}) = \frac{1}{\sqrt{2\pi \hbar}} \mathrm{exp} \bigg [ \frac{i}{\hbar} \bigg ( \vb{K} - \frac{e}{2c} \, \vb{B} \times \vb{r} \bigg ) \vb{R} \bigg ]  \\
{} \times  \mathrm{exp} \bigg ( \frac{i\gamma}{2\hbar} \, \vb{K} \vb{r} \bigg ) \varphi^{(e)}_{nm} (\vb{r} - \vb{r}_0).
\label{eq:eh_wf}
\end{multline}
The corresponding energies involve the cyclotron frequency with the reduced mass, $\omega_0 = |e|B/(\mu c)$, and read
\begin{eqnarray}
E^{(\textup{X},0)}_{nm} &=& \hbar \omega_0 \bigg( \frac{1}{2} + \frac{1+\gamma}{2} \, n + \frac{1-\gamma}{2} \, m \bigg) \nonumber \\
{} &=&  \hbar \omega_0 \bigg( \frac{1}{2} + \tilde{n} + \frac{|\tilde{m}| - \gamma \tilde{m}}{2}  \bigg).
\label{eq:eh_energy}
\end{eqnarray}
Here the superscript ``0'' indicates that the corresponding energies are obtained to zeroth order in the Coulomb interaction. Note that $\omega_0 = \omega_e + \omega_h$, where $\omega_h = |e|B/(m_h c)$. In the heavy-hole limit $m_h = \infty$, one obtains $\gamma = 1$, $\mu = m_e$, $\omega_h = 0$ and recovers the electron LLs~\eqref{eq:1e_energy}.  The angular momentum projection of the two-particle system is again $M_z = -\tilde{m} = n - m$.

Next, one has to compute the matrix elements of the $V$ operator in Eq.~\eqref{eq:H_eh} using the basis functions~\eqref{eq:eh_wf}. The integration over $\vb{R}$ is trivial and the matrix is diagonal with respect to $\vb{K}$:
\begin{multline}\label{eqn:exciton_int_matr_el}
V_{\vb{K}',n'm'; \vb{K},nm} \equiv -(e^2/\lambda) \big \langle \varphi^{(\textup{X})}_{\vb{K}',n'm'} \big | V(r) \big | \varphi^{(\textup{X})}_{\vb{K},nm} \big \rangle \\
= -(e^2/\lambda) \delta (\vb{K} - \vb{K}') \int \dd \vb{r} \big [ \varphi^{(e)}_{n'm'} (\vb{r}) \big ]^* V(|\vb{r} + \vb{r}_0|) \varphi^{(e)}_{nm} (\vb{r}) \\
\equiv \delta (\vb{K} - \vb{K}') V_{\vb{K},n'm'; nm}.
\end{multline}
There are several technical difficulties which one encounters in evaluating the integral over $\vb{r}$. First, note that the function $V(|\vb{r} + \vb{r}_0|)$ is symmetric about $\vb{r} = -\vb{r}_0$, whereas the wave functions are symmetric about the origin (apart from the angular part $\mathrm{e}^{-i\tilde{m}\varphi}$). Second, the radial parts of the wave functions involve special functions (Laguerre polynomials), which complicates the calculations. Moreover, the explicit form of the electrostatic potential may also be highly complex.

In the present study, we compute integrals containing monomials instead of the Laguerre polynomials and then combine the results taking into account the corresponding coefficients. In order to do this, it is necessary to inspect the structure of the wave functions. First, it is convenient to introduce non-normalized wave functions, which will be denoted by $\psi$ (normalized functions are denoted by $\varphi$).  Second, we rewrite the creation operators~\eqref{eqn:cr_an_operators_electron} in terms of a new dimensionless variable $\xi=z/(2\lambda)$:
\begin{align}
 \hat{A}^\dagger_e = \frac{1}{\sqrt{2}} \left(\xi-\pdv{}{\xi^*}\right), \quad \hat{B}^\dagger_e = \frac{1}{\sqrt{2}} \left(\xi^*-\pdv{}{\xi}\right).
\end{align}
The vacuum state~\eqref{eq:vacuum_1el} in terms of $\xi$ reads
\begin{align}
    \Phi^{(e)}_0(\vb{r})=\frac{1}{\lambda \sqrt{2\pi}} \, \mathrm{e}^{-\xi\xi^*}.
\end{align}
The arbitrary wave function can be obtained via
\begin{align}\label{eqn:app_arbitrary_wf}
    \varphi^{(\textup{X})}_{\vb{K},nm} (\vb{r}, \vb{R}) =\phi^{(\textup{X})}_{\vb{K}} (\vb{r}, \vb{R})\dfrac{\psi_{n m}(\vb{r})}{\sqrt{N_{nm}}},
\end{align}
where
\begin{multline}\label{eqn:app_K_part_of_ex_wf}
\phi^{(\textup{X})}_{\vb{K}} (\vb{r}, \vb{R}) = \frac{1}{\sqrt{2\pi \hbar}} \mathrm{exp} \bigg ( \frac{i\gamma}{2\hbar} \, \vb{K} \vb{r} \bigg ) \\
 \times \mathrm{exp} \bigg [ \frac{i}{\hbar} \bigg ( \vb{K} - \frac{e}{2c} \, \vb{B} \times \vb{r} \bigg ) \vb{R} \bigg ],  
\end{multline}
and the relative-motion wave function has the form
\begin{equation}
    \psi_{n m}(\vb{r})= \frac{1}{2^{(n+m)/2}} \bigg (\xi - \pdv{}{\xi^*} \bigg )^n  \bigg ( \xi^* - \pdv{}{\xi} \bigg )^m  \Phi^{(e)}_0(\vb{r}).
\label{eqn:app_psi_nm_operators_explicit}
\end{equation}
The part of Eq.~\eqref{eqn:app_K_part_of_ex_wf} concerning the center-of-mass motion already includes the necessary normalization factor, so $N_{nm}$ should be evaluated as
\begin{align}\label{eqn:app_norm_factor_ex}
N_{nm}=\int \! \dd{\vb{r}} \psi_{n m}^* (\vb{r}) \psi_{n m} (\vb{r}).
\end{align}
Applying operators $\hat{A}^\dagger_e$ and $\hat{B}^\dagger_e$ to the vacuum state according to Eq.~\eqref{eqn:app_psi_nm_operators_explicit}, we obtain the following structure of the wave function $\psi_{nm}$:
\begin{align}\label{eqn:app_gen_wf_e_not_normalized}
    &\psi_{n m}(\vb{r})=\sum_{a_1, a_2} B^{n m}_{a_1 a_2 }(\xi^*)^{a_1}(\xi)^{a_2}\Phi^{(e)}_0(\vb{r}).
\end{align}
The coefficients $B^{n m}_{a_1 a_2 }$ can be found directly by means of Eq.~\eqref{eqn:app_psi_nm_operators_explicit}. We have implemented the corresponding procedure within the \textit{Wolfram Mathematica} subroutine, which is included in the Supplemental Material (for the details see Appendix~\ref{app:math_sr}). 

Taking into account Eq.~\eqref{eqn:app_gen_wf_e_not_normalized}, the normalization factor can be calculated as follows (we do not display the summation indices explicitly):
\begin{multline}\label{eqn:app:normalization_factor_exciton}
N_{nm}=\sum\sum B^{n m*}_{a_1'a_2'}B^{n m}_{a_1 a_2 } \\
\times\int \! \dd{\vb{r}} \big [ \Phi^{(e)}_0(\vb{r}) \big ]^* \Phi^{(e)}_0(\vb{r}) (\xi^*)^{a_1+a_2'}(\xi)^{a_2+a_1'}\\
=\sum\sum B^{n m*}_{a_1'a_2'}B^{n m}_{a_1 a_2} I^{(\textup{X},\textup{N})}_{a_1+a_2',a_2+a_1'}.
\end{multline}
Here we have defined the normalization master integral (MI) $I^{(\textup{X},\textup{N})}_{p_1,p_2}$ involving a monomial:
\begin{widetext}
\begin{align}\label{eqn:app_basic_normalization_ex}
    I^{(\textup{X},\textup{N})}_{p_1,p_2}=\int \! \dd{\vb{r}} \big [ \Phi^{(e)}_0(\vb{r}) \big ]^* \Phi^{(e)}_0(\vb{r}) (\xi^*)^{p_1}(\xi)^{p_2}=\dfrac{1}{2 \pi \lambda^2}\int\limits_0^{\infty}\dd{r} r \mathrm{e}^{-r^2/(2\lambda^2)}\left(\dfrac{r}{2\lambda}\right)^{p_1+p_2}\int\limits_0^{2\pi}\dd{\varphi}\mathrm{e}^{\iu \varphi(p_2-p_1)}=\dfrac{p_1!}{2^{p_1}} \, \delta_{p_1 p_2}.
\end{align}
\end{widetext}
With the aid of the normalization factor~\eqref{eqn:app:normalization_factor_exciton}, one can construct now the normalized wave functions
\begin{align}
    \varphi^{(e)}_{nm} (\vb{r})=\dfrac{\psi_{nm}(\vb{r})}{\sqrt{N_{nm}}},
\end{align}
which coincide with the expression~\eqref{eq:phi_1e_explicit}, and also obtain the full two-particle wave function~\eqref{eqn:app_arbitrary_wf}.

Having described how to calculate the normalization factor $N_{nm}$ using the normalization MI~\eqref{eqn:app_basic_normalization_ex}, we can proceed in the same way and compute the matrix element of the electron-hole interaction $V_{\vb{K},n'm'; nm}$, defined in~Eq.~\eqref{eqn:exciton_int_matr_el}:
\begin{multline}\label{eqn:app_vknm}
V_{\vb{K},n'm'; nm}
=  - \frac{e^2}{\lambda} \int \dd{\vb{r}} \big [ \varphi^{(e)}_{n'm'} (\vb{r}) \big ]^* V(|\vb{r} + \vb{r}_0|) \varphi^{(e)}_{nm} (\vb{r})\\
=\dfrac{\sum\sum B^{n m*}_{a_1'a_2'}B^{n m}_{a_1 a_2} I^{(\textup{X},eh)}_{a_1+a_2',a_2+a_1'}}{\sqrt{N_{n'm'}}\sqrt{N_{nm}}}.
\end{multline}
Here we have introduced the electron-hole interaction MI $I^{(\textup{X},eh)}_{p_1p_2}$:
\begin{widetext}
\begin{multline}\label{eqn:app_basic_interaction_ex}
    I^{(\textup{X},eh)}_{p_1p_2}= - \frac{e^2}{\lambda} \int \! \dd{\vb{r}} \big [ \Phi^{(e)}_0(\vb{r}) \big ]^* \Phi^{(e)}_0(\vb{r})V(|\vb{r}+\vb{r}_0|) (\xi^*)^{p_1}(\xi)^{p_2}= - \dfrac{1}{2\pi}\dfrac{1}{2^{p_1+p_2}} \frac{e^2}{\lambda} \left(\dfrac{r_0}{\lambda}\right)^{2+p_1+p_2}\\ \times\int\limits_0^{\infty}\dd{x}x^{p_1+p_2+1}\mathrm{exp} \bigg [ {-\left(\dfrac{r_0}{\lambda}\right)^2 \dfrac{x^2}{2}}\bigg ] \int\limits_{0}^{2\pi}\dd{\varphi}\mathrm{e}^{\iu \varphi(p_2-p_1)}V(r_0\sqrt{1+x^2+2x\cos{\varphi}}),
\end{multline}
which in the case of the Coulomb potential $V(r) = \lambda/(\varepsilon r)$ reads
\begin{align}\label{eqn:app_basic_interaction_ex_Coulomb}
    I^{(\textup{X},eh,\textup{C})}_{p_1p_2}=-\dfrac{E_0}{\pi\sqrt{2\pi}}\dfrac{1}{2^{p_1+p_2}}\left(\dfrac{r_0}{\lambda}\right)^{1+p_1+p_2}\int\limits_0^{\infty}\dd{x}x^{p_1+p_2+1}\mathrm{exp} \bigg [ {-\left(\dfrac{r_0}{\lambda}\right)^2 \dfrac{x^2}{2}} \bigg ] \int\limits_{0}^{2\pi}\dd{\varphi}\dfrac{\exp[\iu \varphi(p_2-p_1)]}{\sqrt{1+x^2+2x\cos{\varphi}}}.
\end{align}
Here $E_0=\sqrt{\pi/2}(e^2/\varepsilon\lambda)$. With this expression in hands, we are now able to calculate an arbitrary matrix element~\eqref{H_matrix_eh}.

\end{widetext}

\section{Trion (T) in a magnetic field}
\label{sec:app_trinoi_prob}

\subsection{Spectrum calculation recipe}

In contrast to exciton, this three-particle system is not electrically neutral, so one can only partially separate the center-of-mass degrees of freedom (for the details see the next section). Instead of a conserved vector $\vb{K}$, one has in this case an integer quantum number $k$ which can be set to zero due to the degeneracy of the energy levels. The matrix elements read
\begin{align}
H^{(\textup{T})}_{n_1 n_2 n_h m l; n_1' n_2' n_h' m' l'} &= E^{(\textup{T},0)}_{n_1+n_2, n_h} \delta_{n_1 n_1'} \delta_{n_2 n_2'} \delta_{n_h n_h'} \delta_{m m'} \delta_{l l'} \nonumber \\
{} & + V^{(eh)}_{n_1 n_2 n_h m l; n_1' n_2' n_h' m' l'} \nonumber \\
{} &+ V^{(ee)}_{n_1 n_2 n_h m l; n_1' n_2' n_h' m' l'} ,
\label{eq:HT}
\end{align}
where each index is a non-negative integer, $E^{(\textup{T},0)}_{n_e, n_h}$ is given in Eq.~\eqref{eqn:free_LL_trion}. The matrix elements of the interparticle interaction are displayed in Eqs.~\eqref{eqn:app_matr_el_eh} and \eqref{eqn:app_matr_el_ee}. Note that in what follows, we will compute the whole matrix by independently varying all of the indices and taking into account the Coulomb mixing of the free LLs, i.e. incorporating the $V$ terms in Eq.~\eqref{eq:HT_b} with $n_1 \neq n_1'$, $n_2 \neq n_2'$, and $n_h \neq n_h'$.

\subsection{Details}

A three-particle system involving two electrons with positions $\vb{r}_1$ and $\vb{r}_2$ and a hole with position $\vb{r}_h$ is described by the following Hamiltonian:
\begin{equation}
\hat{H}^{(\textup{T})} = \hat{H}_0 + \hat{V}_{eh} + \hat{V}_{ee},
\label{eq:Ham}
\end{equation}
where
\begin{align}
\hat{H}_0 &= \sum_{j=1,2} \frac{\hat{\boldsymbol{\pi}}_j^2}{2m_e} + \frac{\hat{\boldsymbol{\pi}}_h^2}{2m_h}, \\
\hat{V}_{eh} &=  - \sum_{j=1,2} \frac{e^2}{ \lambda} V(|\vb{r}_j - \vb{r}_h|), \\
\hat{V}_{ee} &= \frac{e^2}{ \lambda} V(|\vb{r}_1 - \vb{r}_2|).
\end{align}
Here $\hat{\boldsymbol{\pi}}_j = -i \hbar \boldsymbol{\nabla}_j - (e/c) \boldsymbol{A} (\boldsymbol{r}_j)$. In what follows, we will again construct first the eigenfunctions of the Hamiltonian $\hat{H}_0$ without the interparticle interaction and then use the corresponding wave functions as a basis to subsequently diagonalize the total Hamiltonian $\hat{H}^{(\textup{T})}$.

The most straightforward approach is to employ the one-particle solutions~\eqref{eq:phi_1el} and simply multiply the functions depending on $\vb{r}_1$, $\vb{r}_2$, and $\vb{r}_h$, respectively. However, this basis set completely disregards the translational symmetry of the system as the functions have only a well-defined projection of the total angular momentum $M_z$. First, let us inspect a three-particle generalization of the vector operator~\eqref{eq:K_eh}:
\begin{equation}
\hat{\vb{K}} = \sum_{j=1,2} \hat{\boldsymbol{\pi}}_j + \hat{\boldsymbol{\pi}}_h + \frac{e}{c} \sum_{j=1,2} \vb{B} \times \vb{r}_j - \frac{e}{c} \, \vb{B} \times \vb{r}_h. 
\label{eq:K_eeh}
\end{equation}
This operator (so-called operator of the {\it magnetic translations}~\cite{Lozovikex_1980,PhysRevLett.84.4429, DZYUBENKO2000683, PhysRevB.65.035318}) commutes with both $\hat{H}_0$ and $\hat{H}^{(\textup{T})}$, but the crucial difference between an exciton and trion is that in the latter case the $x$ and $y$ components of $\hat{\vb{K}}$ do not commute with each other. It turns out that $[\hat{K}_x, \, \hat{K}_y ] = i|e|\hbar B /c$ and generally this commutator is proportional to the total electric charge of the system (an exciton is neutral). This means that one can take advantage only of one integral of motion (for instance, either $K_x$ or $K_y$). In fact, it is convenient to consider the operator $\hat{\vb{K}}^2$~\cite{DZYUBENKO2000683}. It obviously commutes with the Hamiltonian. Let us normalize the vector operator according to
\begin{equation}
\hat{\vb{k}} = \sqrt{\frac{c}{|e|\hbar B}} \, \hat{\vb{K}},
\end{equation}
so that $[\hat{k}_x, \, \hat{k}_y] = i$. Introducing then operator $\hat{\varkappa} = (i\hat{k}_x - \hat{k}_y)/\sqrt{2}$, one obtains
\begin{align}
& [\hat{\varkappa}, \, \hat{\varkappa}^\dagger] = 1, \\
& \hat{\vb{K}}^2 = 2 \hat{\varkappa}^\dagger \hat{\varkappa} + 1.
\end{align}
Therefore, the eigenvalues of $\hat{\vb{K}}^2$ are $2k+1$, where $k=0$, $1$, ... The solutions of the Schr\"odinger equation can be characterized by the quantum number $k$, which allows one to partially separate the center-of-mass motion. The question now is how to use this symmetry in practical calculations, i.e., how to construct the eigenfunctions of $\hat{\vb{K}}^2$.

Let us briefly outline the procedure proposed in Ref.~\cite{DZYUBENKO2000683}. First, we perform the following coordinate transformation:
\begin{align}
\vb{R} &= \frac{1}{\sqrt{2}} \, \big ( \vb{r}_1 + \vb{r}_2 \big ), \quad \vb{r} = \frac{1}{\sqrt{2}} \, \big ( \vb{r}_1 - \vb{r}_2 \big ).
\end{align}
Introducing then the three-particle vacuum state
\begin{equation}\label{eqn:old_vacuum}
\Phi^{(\textup{T})}_0 (\vb{r}, \vb{R}, \vb{r}_h) = \Phi^{(e)}_0 (\vb{r}) \Phi^{(e)}_0 (\vb{R}) \Phi^{(e)}_0 (\vb{r}_h),  
\end{equation}
one can generate {\it naive} solutions by the corresponding creation operators of the form~\eqref{eq:Ae} and \eqref{eq:Be} and obtain:
\begin{multline}
\varphi^{(\textup{T})}_{n_1 m_1 n_2 m_2 n_h m_h} (\vb{r}, \vb{R}, \vb{r}_h) \\ 
= \varphi^{(e)}_{n_1 m_1} (\vb{r}) \varphi^{(e)}_{n_2 m_2} (\vb{R}) \varphi^{(h)}_{n_h m_h} (\vb{r}_h),
\label{eq:eeh_naive}
\end{multline}
where $\varphi^{(h)}_{n m} (\vb{r}) = [ \varphi^{(e)}_{n m} (\vb{r}) ]^*$. The $z$ projection of the angular momentum equals $M_z = (n_1 - m_1) + (n_2 - m_2) - (n_h - m_h)$. Moreover, the reflection $\vb{r} \to - \vb{r}$, which is equivalent to exchanging the positions of the electrons, yields the factor $(-1)^{n_1 - m_1}$, so the spin wave function of the electron subsystem (not displayed here explicitly) is even (odd) if $n_1 - m_1$ is odd (even). In other words, the total electron spin is $S_e = 0$ (spin-singlet state) for even $n_1 - m_1$ and $S_e = 1$ (spin-triplet state) for odd $n_1 - m_1$. As was shown in Ref.~\cite{DZYUBENKO2000683}, the functions~\eqref{eq:eeh_naive} can be unitary transformed, so that they become eigenfunctions of $\hat{\vb{K}}^2$, i.e., they gain additional quantum number $k$ preserving $M_z$ and $S_e$. 

To perform this unitary transformation, let us again introduce dimensionless variables $\xi=z/(2\lambda)$, $\xi_R=z_R/(2\lambda)$, and $\xi_h=z_h/(2\lambda)$ and consider the following creation operators:
\begin{align}
\hat{A}^\dag_e(\vb{r})&=\dfrac{\left(\xi-\pdv{}{\xi^*}\right)}{\sqrt{2}}, \ \ \hat{B}^\dag_e(\vb{r})=\dfrac{\left(\xi^*-\pdv{}{\xi}\right)}{\sqrt{2}}, \\
\hat{A}^\dag_e(\vb{R})&=\dfrac{\left(\xi_R-\pdv{}{\xi_R^*}\right)}{\sqrt{2}}, \ \ \hat{B}^\dag_e(\vb{R})=\dfrac{\left(\xi_R^*-\pdv{}{\xi_R}\right)}{\sqrt{2}}, \\
\hat{A}^\dag_h(\vb{r}_h)&=\dfrac{\left(\xi_h^*-\pdv{}{\xi_h}\right)}{\sqrt{2}}, \ \ \hat{B}^\dag_h(\vb{r}_h)=\dfrac{\left(\xi_h-\pdv{}{\xi_h^*}\right)}{\sqrt{2}}.
\end{align}
Let us note that
\begin{equation}
\hat{\varkappa}^\dagger = \sqrt{2}\hat{B}^\dag_e(\vb{R}) - \hat{B}_h(\vb{r}_h).
\end{equation}
This relation can be interpreted as a Bogolyubov transformation of $B^\dag_e(\vb{R})$:
\begin{equation}
  \hat{\widetilde{B}}^\dag_e(\vb{R})=\hat{\varkappa}^\dagger = \sqrt{2}\hat{B}^\dag_e(\vb{R}) - \hat{B}_h(\vb{r}_h) = \hat{S}\hat{B}^\dag_e (\vb{R}) \hat{S}^\dag,\label{eqn:app_Be_tilde}
\end{equation}
The corresponding unitary operator is
\begin{align}
    &\hat{S}=\exp{- \Theta\left [ \hat{B}_e(\vb{R})\hat{B}_h(\vb{r}_h)-\hat{B}^\dag_h(\vb{r}_h)\hat{B}^\dag_e(\vb{R})\right ] },
\end{align}
where $\sinh \Theta = 1$ and $\cosh \Theta=\sqrt{2}$. Another operator that does not commute with $\hat{S}$ and should be transformed is $\hat{B}^\dag_h(\vb{r}_h)$:
\begin{equation}
  \hat{\widetilde{B}}^\dag_h (\vb{r}_h) = \sqrt{2}\hat{B}^\dag_h(\vb{r}_h) - \hat{B}_e(\vb{R}). \label{eqn:app_Bh_tilde}
\end{equation}
In addition to the operators, the wave functions themselves also have to be transformed. Thus, up to the normalization factor the new basis is
\begin{align}
\!\!\!\!    \hat{A}^{\dag}_e(\vb{r})^{n_1}\!\hat{A}^\dag_e(\vb{R})^{n_2}\!\hat{A}^\dag_h(\vb{r}_h)^{n_h}\!\hat{\widetilde{B}}^\dag_e(\vb{R})^k\!\hat{B}^\dag_e(\vb{r})^m\!\hat{\widetilde{B}}_h^\dag(\vb{r}_h)^l \Psi^{(\textup{T})}_0,\!\!\! \label{eqn:app_basis_oper}
\end{align}
where we used a new vacuum state $\Psi^{(\textup{T})}_0 = \hat{S} \Phi^{(\textup{T})}_0$. In this basis, three well-defined 
quantum numbers are $k$, $M_z$, and $S_e$, i.e. the Hamiltonian is block-diagonal in terms of these quantum numbers. The projection of the total angular momentum reads 
\begin{align}
    M_z=\underbrace{n_1-m}_{\textup{from $\vb{r}$}}+\underbrace{n_2-k}_{\textup{from $\vb{R}$}}-\underbrace{(n_h-l)}_{\textup{from $\vb{r}_h$}}.
\end{align}
The spin quantum number is  $S_e = 0$ ($S_e = 1$) if $n_1 - m$ is even (odd). Moreover, hereinafter we will assume $k=0$ since the Hamiltonian $H^{(\textup{T})}$ commutes with $\widetilde{B}^\dag_e(\vb{R})$. The matrix elements do not depend on $k$, so the energy also turns out to be $k$-independent.

In order to obtain an explicit form of the new vacuum state, it is necessary to rewrite the expression for the operator $\hat{S}$. Using the Baker–Campbell–Hausdorff formula (see, e.g., Ref.~\cite{Kirzhnits1967}) for the operator equality $\mathrm{exp} [\Theta(\hat{a}+\hat{b})] = \mathrm{exp} ( \Theta \hat{d} )$, one can present the operator $\hat{S}$ in the following form:
\begin{multline}
        \hat{S}=\mathrm{exp} \left [ \tanh \Theta \hat{B}^\dag_h(\vb{r}_h)\hat{B}^\dag_e(\vb{R}) \right ] \\
        \times\exp{-\ln{(\cosh\Theta)}\!\left[ \hat{B}^\dag_e(\vb{R})\hat{B}_e(\vb{R})+\hat{B}^\dag_h(\vb{r}_h)\hat{B}_h(\vb{r}_h)+1\right ]}\nonumber \\
        \times \mathrm{exp} \left [ -\tanh \Theta \hat{B}_e(\vb{R}) \hat{B}_h(\vb{r}_h) \right ]. \nonumber
\end{multline}
The new vacuum state then reads
\begin{align}
\Psi^{(\textup{T})}_0 =\hat{S} \Phi^{(\textup{T})}_0 =\dfrac{1}{\cosh \Theta}\mathrm{exp} \left [ \tanh \Theta \hat{B}^\dag_h(\vb{r}_h)\hat{B}^\dag_e(\vb{R}) \right ] \Phi^{(\textup{T})}_0.
\end{align}
The state~\eqref{eqn:old_vacuum} is given by
\begin{align}
    \Phi^{(\textup{T})}_0 (\vb{r}, \vb{R}, \vb{r}_h)=\dfrac{\mathrm{exp} (-\xi\xi^*-\xi_R\xi^*_R-\xi_h\xi^*_h ) }{\left(2\pi\right)^{3/2}\lambda^3}.
\end{align}
Taking into account the explicit form of the operators $B^\dag_h(\vb{r}_h)$ and $B^\dag_e(\vb{R})$, one obtains
\begin{multline}
\Psi^{(\textup{T})}_0 = \exp[\dfrac{1}{2\sqrt{2}}\left(\xi_h-\pdv{}{\xi^*_h}\right)\left(\xi^*_R-\pdv{}{\xi_R}\right)]\\
\quad\quad \times\dfrac{\exp(-\xi\xi^*-\xi_R\xi^*_R-\xi_h\xi^*_h)}{\sqrt{2}\left(2\pi\right)^{3/2}\lambda^3}.
\end{multline}
Let us label the function $\sqrt{2} \Psi^{(\textup{T})}_0 (\vb{r}, \vb{R}, \vb{r}_h)$ as a new normalized vacuum state $\tilde{\Phi}^{(\textup{T})}_0 (\vb{r}, \vb{R}, \vb{r}_h)$, which reads
\begin{align}\label{eqn:app_new_vacuun_state}
\tilde{\Phi}^{(\textup{T})}_0\!(\vb{r},\!\vb{R},\vb{r}_h)\!=\!\dfrac{\exp{-\xi\xi^*\!-\!\xi_R\xi^*_R\!-\!\xi_h\xi^*_h\!+\!\sqrt{2}\xi_h\xi^*_R}}{\sqrt{2}\left(2\pi\right)^{3/2}\lambda^3}.\!\!
\end{align}

Now one can generate an arbitrary wave function by means of the creation operators in Eq.~\eqref{eqn:app_basis_oper}. Regarding the normalization of the wave functions, we prefer to keep only factor $1/[\left(2\pi\right)^{3/2}\lambda^3]$ explicitly in contrast to other normalization factors which will be taken into account at the very end. Thus, we first employ the operator $\hat{B}^\dag_e(\vb{r})^m \hat{\widetilde{B}}_h^\dag(\vb{r}_h)^l$. Note that
\begin{align}
    \hat{\widetilde{B}}^\dag_h(\vb{r}_h)^l \tilde{\Phi}^{(\textup{T})}_0\!(\vb{r},\!\vb{R},\vb{r}_h)\!=\dfrac{\left(\xi_h\right)^l\tilde{\Phi}^{(\textup{T})}_0\!(\vb{r},\!\vb{R},\vb{r}_h)\!}{\sqrt{2}}.
\end{align}
Then, we obtain
\begin{multline}
\hat{B}^\dag_e(\vb{r})^m\hat{\widetilde{B}}^\dag_h(\vb{r}_h)^l \tilde{\Phi}^{(\textup{T})}_0\!(\vb{r},\!\vb{R},\vb{r}_h) \\
=2^{(m-1)/2}\left(\xi^*\right)^m\left(\xi_h\right)^l\tilde{\Phi}^{(\textup{T})}_0\!(\vb{r},\!\vb{R},\vb{r}_h).
\end{multline}
It is convenient to introduce the following function:
\begin{align}
  \psi_{ml}(\vb{r},\vb{R},\vb{r}_h)=\left(\xi^*\right)^m\left(\xi_h\right)^l\tilde{\Phi}^{(\textup{T})}_0\!(\vb{r},\!\vb{R},\vb{r}_h).\! 
\end{align}
According to Eq.~\eqref{eqn:app_basis_oper}, the general wave function up to the normalization factor reads
\begin{widetext}
\begin{align}
    \psi_{n_1 n_2 n_h m l}(\vb{r},\vb{R},\vb{r}_h)&=\left[\dfrac{1}{\sqrt{2}}\left(\xi-\pdv{}{\xi^*}\right)\right]^{n_1}\left[\dfrac{1}{\sqrt{2}}\left(\xi_R-\pdv{}{\xi^*_R}\right)\right]^{n_2}\left[\dfrac{1}{\sqrt{2}}\left(\xi^*_h-\pdv{}{\xi_h}\right)\right]^{n_h}\psi_{ml}(\vb{r},\vb{R},\vb{r}_h).
\end{align}
\end{widetext}
Let us now calculate the normalization factor:
\begin{align}\label{eqn:app_norm_factor_tr}
N_{n_1n_2n_hml}=\bothship{\psi_{n_1 n_2 n_h m l}}{\psi_{n_1 n_2 n_h m l}}.
\end{align}
This expression involves three 2D integrals (over $\vb{r}$, $\vb{R}$, and $\vb{r}_h)$. Any wave function can be represented as the following linear combination involving products of $\tilde{\Phi}^{(\textup{T})}_0 (\vb{r},\vb{R},\vb{r}_h)$ and monomials:
\begin{align}\label{eqn:app_gen_wf_not_normalized}
    &\psi_{n_1 n_2 n_h m l}(\vb{r},\vb{R},\vb{r}_h)=\sum C^{n_1 n_2 n_h m l}_{a_1a_2b_1b_2c_1c_2}(\xi^*)^{a_1}(\xi)^{a_2}\nonumber \\
    &\quad\quad \times (\xi_h)^{b_1}(\xi_h^*)^{b_2}(\xi_R)^{c_1}(\xi_R^*)^{c_2}\tilde{\Phi}^{(\textup{T})}_0\!(\vb{r},\!\vb{R},\vb{r}_h).
\end{align}
Therefore,
\begin{widetext}
\begin{align}
  N_{n_1n_2n_hml} &=\sum \sum C^{n_1 n_2 n_h m l *}_{a'_1a'_2b'_1b'_2c'_1c'_2} C^{n_1 n_2 n_h m l}_{a_1a_2b_1b_2c_1c_2} \int \! \dd{\vb{r}} \int \! \dd{\vb{R}}  \int \! \dd{\vb{r}_h} \big | \tilde{\Phi}^{(\textup{T})}_0\!(\vb{r},\!\vb{R},\vb{r}_h) \big |^2 \nonumber\\
  &\qquad\qquad \times (\xi^*)^{a_1+a_2'}(\xi)^{a_2+a_1'}(\xi_h)^{b_1+b_2'}(\xi_h^*)^{b_2+b_1'}(\xi_R)^{c_1+c_2'}(\xi_R^*)^{c_2+c_1'}\nonumber\\
   &=\sum \sum C^{n_1 n_2 n_h m l *}_{a'_1a'_2b'_1b'_2c'_1c'_2} C^{n_1 n_2 n_h m l}_{a_1a_2b_1b_2c_1c_2} I^{(\textup{T},\textup{N})}_{a_1+a_2',a_2+a_1', b_1+ b_2', b_2+b_1',c_1+c_2',c_2+c_1'},
\end{align}
where we define the master integral $I^{(\textup{T},\textup{N})}_{p_1p_2q_1q_2r_1r_2}$ as
\begin{align}
    I^{(\textup{T},\textup{N})}_{p_1p_2q_1q_2r_1r_2} &= \frac{1}{(2 \pi)^3 \lambda^6} \int \! \dd{\vb{r}} \int \! \dd{\vb{R}} \int \! \dd{\vb{r}_h} (\xi^*)^{p_1}(\xi)^{p_2}(\xi_h)^{q_1}(\xi_h^*)^{q_2}(\xi_R)^{r_1}(\xi_R^*)^{r_2} \nonumber \\
    {}&\times \mathrm{exp} \Big [ -2(\xi\xi^*+\xi_R\xi^*_R+\xi_h\xi^*_h)+\sqrt{2}(\xi_h\xi^*_R+\xi_h^*\xi_R) \Big ].
    \label{eqn:app_i_c_basic_integral}
\end{align}
\end{widetext}
The expansion coefficients $C^{n_1 n_2 n_h m l}_{a_1a_2b_1b_2c_1c_2}$ for an arbitrary wave function in terms of monomials $(\xi^*)^{a_1}(\xi)^{a_2}(\xi_h)^{b_1}(\xi_h^*)^{b_2}(\xi_R)^{c_1}(\xi_R^*)^{c_2}$ can be obtained automatically. We attach the corresponding subroutine in \textit{Wolfram Mathematica} format. Thus, we should evaluate only the master integral~\eqref{eqn:app_i_c_basic_integral}. First, taking into account $\xi=r \mathrm{e}^{\iu \varphi} /(2\lambda)$, we integrate out the $\vb{r}$-dependent part:
\begin{widetext}
\begin{align}
\int\dd{\vb{r}}\left(\xi^*\right)^{p_1}\left(\xi\right)^{p_2}\nonumber \exp(-2\xi\xi^*)=\int\limits^\infty_0r\dd{r}\int\limits^{2\pi}_0\dd{\varphi} \exp[\iu\varphi(p_2-p_1)]\left(\dfrac{r}{2\lambda}\right)^{p_1+p_2} \exp(-\dfrac{r^2}{2\lambda^2})=2\pi\delta_{p_1p_2}\dfrac{p_1!\lambda^2}{2^{p_1}}.
\end{align}
Equation~\eqref{eqn:app_i_c_basic_integral} takes the form
\begin{multline}
I^{(\textup{T},\textup{N})}_{p_1p_2q_1q_2r_1r_2}=\dfrac{2\pi\delta_{p_1p_2}p_1!}{(2\pi)^3 2^{p_1}\lambda^4}\int\limits^\infty_0R\dd{R}\int\limits^\infty_0r_h\dd{r_h}\left(\dfrac{r_h}{2\lambda}\right)^{q_1+q_2}\left(\dfrac{R}{2\lambda}\right)^{r_1+r_2}\int\limits_{0}^{2\pi}\dd{\varphi_h}\int\limits_{0}^{2\pi}\dd{\varphi_R}
    \exp[\iu\varphi_h(q_1-q_2)]\exp[\iu\varphi_R(r_1-r_2)] \\
     \times\exp[-\dfrac{R^2}{2\lambda^2}-\dfrac{r_h^2}{2\lambda^2}+\dfrac{r_hR\cos{(\varphi_R-\varphi_h)}}{\sqrt{2}\lambda^2}].
\end{multline}
Using the Jacobi-Anger relation $\mathrm{exp} (\iu z\cos{\varphi}) = \!\!\!\! \sum\limits_{n=-\infty}^{\infty}\!\!\!\!\iu^nJ_n(z)\mathrm{exp} (\iu n \varphi)$, we can integrate over the angles as follows:
\begin{align}
&\int\limits_0^{2\pi}\dd{\varphi_R}\int\limits_0^{2\pi}\dd{\varphi_h}\exp[\iu\varphi_h(q_1-q_2)]\exp[\iu\varphi_R(r_1-r_2)]\exp[\iu n(\varphi_R-\varphi_h)]=(2\pi)^2 \delta_{0,q_1-q_2-n}\delta_{0,r_1-r_2+n}.
\end{align}
Let us assume that $q_1\geq q_2$. One obtains
\begin{align}
I^{(\textup{T},\textup{N})}_{p_1p_2q_1q_2r_1r_2}&=\frac{1}{\lambda^4} \, \delta_{p_1p_2}\delta_{q_1-q_2,r_2-r_1} \, \dfrac{p_1!}{2^{p_1}} \, \iu^{q_1-q_2}\int\limits^\infty_0\dd{R} R \left(\dfrac{R}{2\lambda}\right)^{r_1+r_2}\exp(-\dfrac{R^2}{2\lambda^2})\nonumber \\
&\times \int\limits^\infty_0\dd{r_h} r_h \left(\dfrac{r_h}{2\lambda}\right)^{q_1+q_2} \exp(-\dfrac{r_h}{2\lambda^2}) J_{q_1-q_2}\left(-\iu\dfrac{r_hR}{\lambda^2\sqrt{2}}\right) =\dfrac{\delta_{p_1p_2}\delta_{q_1-q_2,r_2-r_1}p_1!}{2^{p_1+q_1+r_1+\frac{q_1-q_2}{2}-2}}\int\limits^\infty_0\dd{\left(\dfrac{R}{\sqrt{2}\lambda}\right)}\left(\dfrac{R}{\sqrt{2}\lambda}\right) \nonumber\\
    &\times \int\limits^\infty_0\dd{\left(\dfrac{r_h}{\sqrt{2}\lambda}\right)}\left(\dfrac{r_h}{\sqrt{2}\lambda}\right) \sum\limits^\infty_{m=0}\dfrac{1}{2^m m!}\dfrac{1}{(q_1-q_2+m)!}\exp(-\dfrac{r^2_h}{2\lambda^2})\exp(-\dfrac{R^2}{2\lambda^2}) \left(\dfrac{r_h}{\sqrt{2}\lambda}\right)^{2q_1+2m} \nonumber\\
    &\times \left(\dfrac{R}{\sqrt{2}\lambda}\right)^{2r_1+2m+2q_1-2q_2} =\dfrac{\delta_{p_1p_2}\delta_{q_1-q_2,r_2-r_1}p_1!}{2^{p_1+q_1+r_1+\frac{q_1-q_2}{2}}}\int\limits^\infty_0 \dd{x}\int\limits^\infty_0 \dd{y} \sum\limits^\infty_{m=0}\dfrac{y^{q_1+m}\exp(-y)x^{m+r_1+q_1-q_2}\exp(-x)}{2^m m!\left(q_1-q_2+m\right)!}=\nonumber\\
    &=\dfrac{\delta_{p_1p_2}\delta_{q_1-q_2,r_2-r_1}p_1!}{2^{p_1+q_1+r_1+\frac{q_1-q_2}{2}}} \sum\limits_{m=0}^{\infty} \dfrac{\left(q_1+m\right)!(r_1+m+q_1-q_2)!}{m!2^m\left(q_1-q_2+m\right)!}=\dfrac{\delta_{p_1p_2}\delta_{q_1-q_2,r_2-r_1}p_1!q_1!r_2!}{(q_1-q_2)!2^{p_1+q_1+r_1+\frac{q_1-q_2}{2}}} \nonumber \\
    &\times {}_2F_1 \left(q_1 + 1 ,  r_2  + 1 , q_1 - q_2  +1 ; \dfrac{1}{2}\right).
\end{align}
\end{widetext}
If $q_1 < q_2$, then one can perform complex conjugation of the initial expression for $I^{(\textup{T},\textup{N})}_{p_1p_2q_1q_2r_1r_2}$ and calculate it via the same steps as described above. The result can also be obtained by simply substituting $q_1\leftrightarrow q_2$ and $r_1\leftrightarrow r_2$. Then, for universality, we introduce $q_{>}=\textup{max}(q_1,q_2)$, $r_{>}=\textup{max}(r_1,r_2)$, and $r_{<}=\textup{min}(r_1,r_2)$. In terms of these quantities, the master integral reads
\begin{widetext}
\begin{align}
    I^{(\textup{T},\textup{N})}_{p_1p_2q_1q_2r_1r_2}=\!\dfrac{\delta_{p_1p_2}\delta_{q_1-q_2,r_2-r_1}p_1!q_{>}!r_{>}!}{|q_1-q_2|!2^{p_1+q_{>}+r_{<}+\frac{|q_1-q_2|}{2}}}\! \ _2F_1\!\!\left(\!q_> \! + \! 1 ,  r_> \! +\! 1 , |q_1\! - \! q_2| \! +1 ; \!\dfrac{1}{2}\!\right).
\end{align}
\end{widetext}
Having obtained the expression for the normalization factor $N_{n_1n_2n_hml}$, we introduce the normalized wave functions as 
\begin{align}
\varphi^{(\textup{T})}_{n_1 n_2 n_h m l}(\vb{r},\vb{R},\vb{r}_h)=\dfrac{\psi_{n_1 n_2 n_h m l}(\vb{r},\vb{R},\vb{r}_h)}{\sqrt{N_{n_1n_2n_hml}}}.
\end{align}
With these wave functions, we can calculate the matrix elements of interest. First, it immediately follows that the matrix elements of the operator $\hat{H}_0$ read
\begin{multline}\label{eqn:app_free_energy}
\tripleship{\varphi^{(\textup{T})}_{n_{1} n_{2} n_{h} m l}}{\hat{H}_0}{\varphi^{(\textup{T})}_{n_{1}' n_{2}' n_{h}' m' l'}}\\ =\delta_{n_{1}n_{1}'}\delta_{n_{2}n_{2}'}\delta_{n_{h}n_{h}'}\delta_{m m'}\delta_{l l'}E^{(\textup{T},0)}_{n_1+n_2, n_h},
\end{multline}
where
\begin{equation}\label{eqn:free_LL_trion}
E^{(\textup{T},0)}_{n_e n_h} = \hbar \omega_e \bigg( 1 + n_e \bigg) + \hbar \omega_h \bigg( \frac{1}{2} + n_h \bigg).
\end{equation}

Next, one has to compute the matrix elements involving the interparticle-interaction terms. Let us start with the contribution $\hat{V}_{eh}$. Taking into account Eq.~\eqref{eqn:app_gen_wf_not_normalized}, we obtain
\begin{widetext}
\begin{align}\label{eqn:app_matr_el_eh}
  V^{(eh)}_{n_1 n_2 n_h m l; n_1' n_2' n_h' m' l'} \!\!=\!\!\tripleship{\varphi^{(\textup{T})}_{n_{1} n_{2} n_{h} m l}}{\hat{V}_{eh}}{\varphi^{(\textup{T})}_{n_{1}' n_{2}' n_{h}' m' l'}}\!=\!\!\dfrac{\sum \sum C^{n_1 n_2 n_h m l *}_{a'_1a'_2b'_1b'_2c'_1c'_2} C^{n_1 n_2 n_h m l}_{a_1a_2b_1b_2c_1c_2} I^{(\textup{T},eh)}_{a_1+a_2',a_2+a_1', b_1+ b_2', b_2+b_1',c_1+c_2',c_2+c_1'}}{\sqrt{N_{n_1n_2n_hml}}\sqrt{N_{n_1'n_2'n_h'm'l'}}}, 
\end{align}
where 
\begin{multline}
I^{(\textup{T},eh)}_{p_1p_2q_1q_2r_1r_2}=-\dfrac{e^2}{\lambda} \!\int \!\! \dd{\vb{r}}\!\! \int \!\! \dd{\vb{R}} \!\!\int \!\! \dd{\vb{r}}_h(\xi^*)^{p_1}(\xi)^{p_2}(\xi_h)^{q_1}(\xi_h^*)^{q_2}(\xi_R)^{r_1}(\xi_R^*)^{r_2} \\ \times \dfrac{\exp{-2(\xi\xi^*+\xi_R\xi^*_R+\xi_h\xi^*_h)+\sqrt{2}(\xi_h\xi^*_R+\xi_h^*\xi_R)}}{\left(2\pi\right)^{3}\lambda^6}\left[V(|\vb{r}_h-\vb{r}_1|)+V(|\vb{r}_h-\vb{r}_2|)\right].
\end{multline}
It is convenient to introduce the variable $\widetilde{\vb{r}}=\sqrt{2}\vb{r}_h-\vb{R}$. We perform further calculations for an arbitrary spherically symmetric potential, while the Coulomb potential will be considered only at the very end of our calculations. In terms of $\widetilde{\vb{r}}$ , the sum in square brackets can be written as $V(|\widetilde{\vb{r}}-\vb{r}|/\sqrt{2})+V(|\widetilde{\vb{r}}+\vb{r}|/2)$. Passing to the coordinates $\{\widetilde{\vb{r}},\vb{r},\vb{R}\}$ and also making an additional change of variables in the second term [reflection $\vb{r}\to  -\vb{r}$ leading to the factor $(-1)^{p_1+p_2}$], we obtain the following expression:
\begin{align}    
    I^{(\textup{T},eh)}_{p_1p_2q_1q_2r_1r_2}&=-\dfrac{e^2[1+(-1)^{p_1+p_2}]}{2\lambda(2\pi)^3\lambda^6} \int\dd{\vb{r}}\int\dd{\vb{R}}\int\dd{\widetilde{\vb{r}}}(\xi)^{p_2}(\xi^*)^{p_1}(\xi_R)^{r_1}(\xi_R^*)^{r_2}\left(\dfrac{\widetilde{\xi}+\xi_R}{\sqrt{2}}\right)^{q_1}\left(\dfrac{\widetilde{\xi}^*+\xi^*_R}{\sqrt{2}}\right)^{q_2}\nonumber\\
    &\times V\left(\dfrac{|\widetilde{\vb{r}}-\vb{r}|}{\sqrt{2}}\right)\exp(-\dfrac{r^2}{2\lambda^2}-\dfrac{\widetilde{r}^2}{4\lambda^2}-\dfrac{R^2}{4\lambda^2})=-\dfrac{e^2(1+(-1)^{p_1+p_2})}{2\lambda (2\pi)^3\lambda^62^{(q_1+q_2)/2}}\sum\limits_{k=0}^{q_1}\sum\limits_{n=0}^{q_2}C^k_{q_1}C^n_{q_2}\int\dd{\vb{r}}\int\dd{\widetilde{\vb{r}}} \nonumber\\
    &\times V\left(\dfrac{|\widetilde{\vb{r}}-\vb{r}|}{\sqrt{2}}\right)\exp(-\dfrac{r^2}{2\lambda^2}-\dfrac{\widetilde{r}^2}{4\lambda^2})(\xi)^{p_2}(\xi^*)^{p_1}(\widetilde{\xi})^{q_1-k}(\widetilde{\xi}^*)^{q_2-n}\int\dd{\vb{R}}(\xi_R)^{r_1+k}(\xi^*_R)^{r_2+n}\exp(-\dfrac{R^2}{4\lambda^2}),
    \label{eqn:app_ieh_exp}
\end{align}
where $\widetilde{\xi}=\sqrt{2}\xi_h-\xi_R$. Let us separately consider $\vb{R}$ and $\widetilde{\vb{r}}$--$\vb{r}$ parts. For the $\vb{R}$ integration, we obtain the following expression:
\begin{align}
    &\int\limits^\infty_0\dd{R} R \exp(-\dfrac{R^2}{4\lambda^2})\left(\dfrac{R}{2\lambda}\right)^{r_1+r_2+k+n}\int\limits^{2\pi}_0\dd{\varphi_R}\!\exp{\iu\varphi_R[r_1+k-(r_2+n)]}=4\pi\lambda^2\delta_{r_1+k,r_2+n}(k+r_1)!  .
\end{align}
Now we consider the $\widetilde{\vb{r}}$--$\vb{r}$ integration in Eq.~\eqref{eqn:app_ieh_exp}. In order to factorize the expression, it suffices here to perform the following change of variables: $\vb{r}=\bm{\rho}_1/3+\bm{\rho}_2/3$, $\widetilde{\vb{r}}=\bm{\rho}_1/3-2\bm{\rho}_2/3$.  Then, the corresponding integral reads
\begin{align}
    &\int\dd{\vb{r}}\int\dd{\widetilde{\vb{r}}}V\left(\dfrac{|\widetilde{\vb{r}}-\vb{r}|}{\sqrt{2}}\right)\exp(-\dfrac{r^2}{2\lambda^2}-\dfrac{\widetilde{r}^2}{4\lambda^2})(\xi)^{p_2}(\xi^*)^{p_1}(\widetilde\xi)^{q_1-k}(\widetilde\xi^*)^{q_2-n}=\dfrac{1}{3^{2+p_2+p_1+q_1+q_2-k-n}}\int\dd\bm{\rho}_1\!\!\int\dd\bm\rho_2V\!\!\left(\dfrac{|\bm{\rho}_2|}{\sqrt{2}}\right)\nonumber\\ 
    &\times\exp(-\dfrac{\rho_1^2+2\rho_2^2}{12\lambda^2})\left(\xi_{\rho_1}+\xi_{\rho_2}\right)^{p_2}\left(\xi^*_{\rho_1}+\xi_{\rho_2}^*\right)^{p_1}\left(\xi_{\rho_1}-2\xi_{\rho_2}\right)^{q_1-k}\left(\xi^*_{\rho_1}-2\xi^*_{\rho_2}\right)^{q_2-n}=\dfrac{1}{3^{2+p_2+p_1+q_1+q_2-k-n}}\nonumber\\ &\times\sum\limits_{t_2=0}^{p_2}\sum\limits_{t_1=0}^{p_1}\sum\limits_{s_1=0}^{q_1-k}\sum\limits_{s_2=0}^{q_2-n}C^{t_2}_{p_2}C^{t_1}_{p_1}C^{s_1}_{q_1-k}C^{s_2}_{q_2-n}\int\limits_{0}^\infty\rho_2\dd{\rho_2}V\left(\dfrac{\rho_2}{\sqrt{2}}\right)\exp(-\dfrac{\rho_2^2}{6\lambda^2})(\xi_{\rho_2})^{p_2-t_2}(\xi^*_{\rho_2})^{p_1-t_1}(-2\xi_{\rho_2})^{q_1-l-s_1}\nonumber\\&\times(-2\xi^*_{\rho_2})^{q_2-n-s_2}\int\rho_1\dd{\rho_1}(\xi_{\rho_1})^{t_2+s_1}(\xi_{\rho_1}^*)^{t_1+s_2}\exp(-\dfrac{\rho^2_1}{12\lambda^2}) =(*).\label{eqn:app_r_rt_part}
\end{align}
The $\rho_1$- and $\rho_2$-integrals are factorized, so they can be easily evaluated separately. First,
\begin{align}
    &\int\limits_{0}^{\infty}\dd{\rho_1}\rho_1\!\left(\dfrac{\rho_1}{2\lambda}\right)^{t_2+t_1+s_1+s_2}\!\!\!\!\!\exp(-\dfrac{\rho_1^2}{12\lambda^2})\!\!\int\limits^{2\pi}_{0}\!\dd{\varphi_{\rho_1}}\!\exp[\iu\varphi_{\rho_1}(t_2\!+\!s_1\!-\!(t_1\!+\!s_2))]=12\pi\lambda^23^{t_1+s_2}(t_1+s_2)!\delta_{t_2+s_1, t_1+s_2}.
\end{align}
Second,
\begin{align}
    &\int\limits_0^\infty \dd{\rho_2} \rho_2V\left(\dfrac{\rho_2}{\sqrt{2}}\right)\left(\dfrac{\rho_2}{2\lambda}\right)^{p_2-t_2+p_1-t_1+q_1-k-s_1+q_2-n-s_2}\exp(-\dfrac{\rho_2^2}{6\lambda^2})\int\limits^{2\pi}_{0}\dd{\varphi_{\rho_2}}(-2)^{q_1-k-s_1+q_2-n-s_2}\nonumber\\ 
    &\qquad\qquad\qquad\qquad\qquad\qquad\qquad\qquad\times\exp[\iu\varphi_{\rho_2}(p_2-t_1+q_1-k-s_1-(p_1-t_1+q_2-n-s_2))]
    =\nonumber\\
    &=2\pi\delta_{p_2+q_1-k,p_1+q_2-n}(-2)^{q_1-k-s_1+q_2-n-s_2}\int\limits_0^\infty \dd{\rho_2} V\!\left(\dfrac{\rho_2}{\sqrt{2}}\right)\rho_2\exp(-\dfrac{\rho_2^2}{6\lambda^2})\left(\dfrac{\rho_2^2}{4\lambda^2}\right)^{p_2-t_2+q_1-k-s_1}=\nonumber\\
    &=2\pi\delta_{p_2+q_1-k,p_1+q_2-n}(-2)^{q_1-k-s_1+q_2-n-s_2} \lambda^2 \mathfrak{I}_{eh}(p_2-t_2+q_1-k-s_1),
\end{align}
where we have introduced
\begin{align}\label{app_eqn:general_radial_int_trion}
    \mathfrak{I}_{eh}(\alpha)=\dfrac{1}{\lambda^2}\int\limits_0^\infty V\left(\dfrac{\rho_2}{\sqrt{2}}\right)\exp(-\dfrac{\rho_2^2}{6\lambda^2})\left(\dfrac{\rho_2^2}{4\lambda^2}\right)^{\alpha} \rho_2\dd{\rho_2}.
\end{align}
In the case of Coulomb potential, this integral yields
\begin{align}
    \mathfrak{I}^{\textup{C}}_{eh}(\alpha)=\dfrac{\sqrt{3\pi}}{\varepsilon}\left(\dfrac{3}{4}\right)^{\alpha}\left ( 2\alpha-1\right )!!. 
\end{align}
Collecting the contributions for the $\widetilde{\vb{r}}$--$\vb{r}$ part, for the expression~\eqref{eqn:app_r_rt_part}, we finally obtain
\begin{multline}
    (*)=\dfrac{1}{3^{2+p_2+p_1+q_1+q_2-k-n}}\sum\limits_{t_2=0}^{p_2}\sum\limits_{t_1=0}^{p_1}\sum\limits_{s_1=0}^{q_1-k}\sum\limits_{s_2=0}^{q_2-n}C^{t_2}_{p_2}C^{t_1}_{p_1}C^{s_1}_{q_1-k}C^{s_2}_{q_2-n}12\pi \lambda^23^{t_1+s_2}(t_1+s_2)!\delta_{t_2+s_1,t_1+s_2}\\
    \times 2\pi\delta_{p_2+q_1-k,p_1+q_2-n}(-2)^{q_1-k-s_1+q_2-n-s_2} \lambda^2 \mathfrak{I}^{C}_{eh}(p_2-t_2+q_1-k-s_1).
\end{multline}
Thus, we find the final expression for $I^{(\textup{T},eh)}_{p_1p_2q_1q_2r_1r_2}$. Continuing the equation~\eqref{eqn:app_ieh_exp}, we come to the following expression:
\begin{align}
I^{(\textup{T},eh)}_{p_1p_2q_1q_2r_1r_2}=&-\delta_{p_2+q_1+r_1,p_1+q_2+r_2}\dfrac{2e^2}{3\lambda}
\dfrac{(1+(-1)^{p_1+p_2})}{2^{(q_1+q_2)/2}}\sum\limits_{k=0}^{q_1}\sum\limits_{n=0}^{q_2}C^k_{q_1}C^n_{q_2}\dfrac{\delta_{r_1+k,r_2+n}(k+r_1)!}{3^{p_1+p_2+q_1+q_2-k-n}}\nonumber\\
&\times\sum\limits_{t_2=0}^{p_2}\sum\limits_{t_1=0}^{p_1}\sum\limits_{s_1=0}^{q_1-k}\sum\limits_{s_2=0}^{q_2-n}C^{t_2}_{p_2}C^{t_1}_{p_1}C^{s_1}_{q_1-k}C^{s_2}_{q_2-n} 3^{t_1+s_2}(t_1+s_2)!\delta_{t_2+s_1,t_1+s_2}(-2)^{q_1-k-s_1+q_2-n-s_2} \nonumber \\
&{}\times \mathfrak{I}_{eh}(p_2-t_2+q_1-k-s_1).
\end{align}
This is the most general expression allowing one to obtain the matrix element~\eqref{eqn:app_matr_el_eh} for any spherically symmetric potential.
In the case of the Coulomb one, it takes the following form:
\begin{align}
    &I^{(\textup{T},eh,\textup{C})}_{p_1p_2q_1q_2r_1r_2}=-\delta_{p_2+q_1+r_1,p_1+q_2+r_2}\dfrac{2\sqrt{2}E_0}{\sqrt{3}}\dfrac{(-1)^{q_1+q_2}[1+(-1)^{p_1+p_2}]}{3^{p_1+q_2}2^{p_1+p_2+(q_1+q_2)/2}}\sum\limits_{k=0}^{q_1}\sum\limits_{n=0}^{q_2}C^k_{q_1}C^n_{q_2}\delta_{r_1+k,r_2+n}(k+r_1)!\\
    &\times\sum\limits_{t_2=0}^{p_2}\sum\limits_{t_1=0}^{p_1}\sum\limits_{s_1=0}^{q_1-k}\sum\limits_{s_2=0}^{q_2-n}C^{t_2}_{p_2}C^{t_1}_{p_1}C^{s_1}_{q_1-k}C^{s_2}_{q_2-n}\delta_{t_2+s_1,t_1+s_2}(-1)^{k+s_1+n+s_2}3^{t_1+s_2+n}2^{t_1+t_2}(t_1+s_2)!\!\left[2(p_2\!-\!t_2\!+\!q_1\!-\!k\!-\!s_1\!)\!-\!1\!\right]!!.\nonumber
\end{align}

The matrix elements corresponding to the electron-electron interaction is determined via
\begin{align}\label{eqn:app_matr_el_ee}
  V^{(ee)}_{n_{1} n_{2} n_{h} m l n_{1}' n_{2}' n_{h}' m' l'} \!\!=\!\!\tripleship{\varphi^{(\textup{T})}_{n_1 n_2 n_h m l}}{\hat{V}_{ee}}{\varphi^{(\textup{T})}_{n_{1}' n_{2}' n_{h}' m' l'}}\!=\!\!\dfrac{\sum \sum C^{n_1 n_2 n_h m l *}_{a'_1a'_2b'_1b'_2c'_1c'_2} C^{n_1 n_2 n_h m l}_{a_1a_2b_1b_2c_1c_2} I^{(\textup{T},ee)}_{a_1+a_2',a_2+a_1', b_1+ b_2', b_2+b_1',c_1+c_2',c_2+c_1'}}{\sqrt{N_{n_1n_2n_hml}}\sqrt{N_{n_1'n_2'n_h'm'l'}}}, 
\end{align}
where 
\begin{multline}
I^{(\textup{T},ee)}_{p_1p_2q_1q_2r_1r_2}=-\dfrac{e^2}{\lambda} \!\int \!\! d\vb{r}\!\! \int \!\! d\vb{R} \!\!\int \!\! d\vb{r}_h(\xi^*)^{p_1}(\xi)^{p_2}(\xi_h)^{q_1}(\xi_h^*)^{q_2}(\xi_R)^{r_1}(\xi_R^*)^{r_2} \\ \times \dfrac{\exp[-2(\xi\xi^*+\xi_R\xi^*_R+\xi_h\xi^*_h)+\sqrt{2}(\xi_h\xi^*_R+\xi_h^*\xi_R)]}{\left(2\pi\right)^{3}\lambda^6}V(|\vb{r}_2-\vb{r}_1|).
\end{multline}
One then obtains
\begin{align}
    I^{(\textup{T},ee)}_{p_1p_2q_1q_2r_1r_2}&=\dfrac{1}{(2\pi)^3\lambda^6}\dfrac{e^2}{\lambda}\int\limits_0^{\infty}r\dd{r}\int\limits_0^{2\pi}\dd{\varphi}V(\sqrt{2}r)\left(\dfrac{r}{2\lambda}\right)^{p1+p2}\exp(-\dfrac{r^2}{2\lambda^2})\exp[\iu \varphi(p_2-p_1)]\int\limits r_h \dd{r_h}\left(\dfrac{r_h}{2\lambda}\right)^{q_1+q_2}\nonumber\\
    &\times\exp(-\dfrac{r_h^2}{2\lambda^2})\int\limits_{0}^{\infty}\!\dd{R} R\left(\dfrac{R}{2\lambda}\right)^{r_1+r_2}\exp(-\dfrac{R^2}{2\lambda^2})\int\limits_0^{2\pi}\dd{\varphi_R}\int\limits_0^{\infty}\dd{\varphi_{h}}\exp[\iu \varphi_R(r_1-r_2)]\exp[\iu \varphi_{h}(q_1-q_2)]\nonumber\\
    &\times\exp[\iu u\cos{(\varphi_R-\varphi_{h})}],
\end{align}
where $u=-\iu R r_h/(\sqrt{2}\lambda^2)$. The integration over the angles yields
\begin{multline}
\int\limits_0^{2\pi}\dd{\varphi_{r}}\exp[\iu \varphi_r(p_2-p_1)]\int\limits_0^{2\pi}\dd{\varphi_R}\int\limits_0^{\infty}\dd{\varphi_{r_{h}}}\exp[\iu \varphi_R(r_1-r_2)]\exp[\iu \varphi_{r_{h}}(q_1-q_2)]\exp[\iu u\cos{(\varphi_R-\varphi_{r_h})}]\\
=\delta_{p_1p_2}\delta_{q_1-q_2,r_1-r_2}\iu ^{r_2-r_1}(2\pi)^3J_{r_2-r_1}(u).
\end{multline}
Expanding the Bessel function in a power series and integrating over $R$ and $r_h$, we obtain the final expression for an arbitrary spherically symmetric potential:
\begin{align}
    I^{(\textup{T},ee)}_{p_1p_2q_1q_2r_1r_2}=\dfrac{e^2}{\lambda}\dfrac{1}{2^{q_2+(r_2+r_1)/2}}\delta_{p_1p_2}\delta_{r_2-r_1,q_1-q_2}\mathfrak{I}_{ee}(2p_1)\sum\limits_{m=0}^{\infty}\dfrac{1}{2^m m!}\dfrac{(m+q_1)!(m+r_2)!}{(m+r_2-r_1)!},
\end{align}
where we define the $r$ part via

\begin{align}
    \mathfrak{I}_{ee}(\alpha)=\dfrac{1}{\lambda^2}\int\limits_0^{\infty}r\dd{r}V(\sqrt{2}r)\left(\dfrac{r}{2\lambda}\right)^{\alpha} \exp(-\dfrac{r^2}{2\lambda^2}).
\end{align}
In the case of the Coulomb potential, this integral is given by
\begin{align}
    \mathfrak{I}_{ee}^{\textup{C}}(\alpha)=\dfrac{\sqrt{\pi}}{2\varepsilon}\dfrac{\left(\alpha-1\right)!!}{2^{\alpha}}.
\end{align}
Thus, for the Coulomb potential, the master integral of the electron-electron interaction reads   
\begin{align}
    I^{(\textup{T},ee,\textup{C})}_{p_1p_2q_1q_2r_1r_2}=\dfrac{E_0}{\sqrt{2}}\dfrac{\left(2p_1-1\right)!!}{2^{q_2+(r_2+r_1)/2+2p_1}}\delta_{p_1p_2}\delta_{r_2-r_1,q_1-q_2}\sum\limits_{m=0}^{\infty}\dfrac{1}{2^m m!}\dfrac{(m+q_1)!(m+r_2)!}{(m+r_2-r_1)!}.
\end{align}
\end{widetext}
Now we have all of the necessary ingredients for the Hamiltonian matrix elements~\eqref{eq:HT}.

\section{Exciton and a free electron $(\textup{X}, e)$} \label{sec:X_e}
When solving the three-particle problem discussed in Appendix~\ref{sec:app_trinoi_prob}, one obtains discrete energy levels corresponding to bound trion states and a continuous spectrum. The lower boundary of the latter indicates the energy at which the trion disintegrates and turns into a combination of an exciton and a free electron. Since these two constituents do not interact with each other, the free electron contributes to the total energy according to Eq.~\eqref{eq:1e_energy}, while the exciton contribution can be extracted from the matrix problem~\eqref{H_matrix_eh}. Let the free-electron LL be characterized by quantum number $n_0$. The Hamiltonian is diagonal with respect to $n_0$ because the Coulomb interaction relates solely to the exciton part. The matrix has the following form:
\begin{multline}
H^{(\textup{X},e)}_{\vb{K},n_0,n'm'; nm} = E^{(\textup{X},0)}_{nm} \delta_{nn'} \delta_{mm'} \\+ V_{\vb{K}, n'-n_0,m';n-n_0,m}.
\label{eq:H_matrix_X_e}
\end{multline}
Here the second term is present only for $n_0 \leqslant \mathrm{min} \, (n, \, n')$. To obtain the lower energy level of the continuous spectrum, one also has to minimize the resulting energy by varying $|\vb{K}|$. If the Coulomb mixing among different LLs of the three-particle problem is neglected, then for given $n_e \equiv n_1 + n_2$ and $n_h$ in Eq.~\eqref{eq:HT} one should choose $n = n' = n_e$ and $m = m' = n_h$ in Eq.~\eqref{eq:H_matrix_X_e}. In order to fully take into account the Coulomb interaction, it is necessary to vary $n$, $n'$, $m$, $m'$ in Eq.~\eqref{eq:H_matrix_X_e} from zero to sufficiently large numbers, so the energy eigenvalues converge.

\section{Examples of calculating the continuum onset in the simplest cases}
\label{sec:app_cont_onset}

\begin{figure}[t!]
    \centering
    \includegraphics[width = 1.0\linewidth]{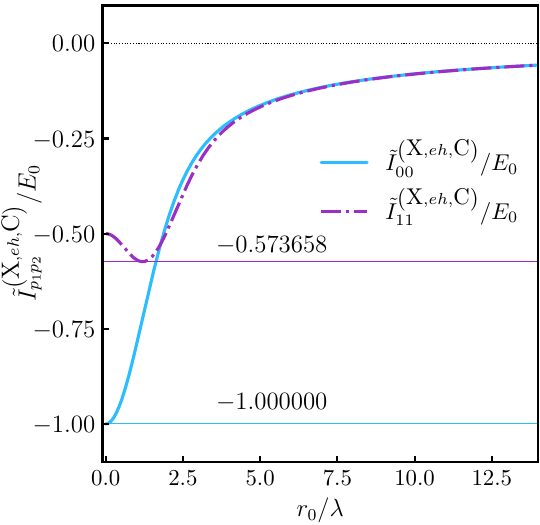}
    \caption{
    Functions $\tilde{I}^{(\textup{X},eh,\textup{C})}_{00}=I^{(\textup{X},eh,\textup{C})}_{00}$ and  $\tilde{I}^{(\textup{X},eh,\textup{C})}_{11}=2I^{(\textup{X},eh,\textup{C})}_{11}$ governing the Coulomb matrix elements and defined in Eq.~\eqref{eqn:app_basic_interaction_ex_Coulomb} versus $r_0 = |\vb{r}_0|$ in units of $\lambda$. These quantities entering Eqs.~\eqref{eqn:ex_e_ex_min_00} and~\eqref{eqn:ex_e_ex_min_01} determine the exciton energy. The vector $\vb{r}_0$ is defined via $\vb{r}_0 = -c(\vb{B} \times \vb{K})/(eB^2)$, where $\vb{K}$ is a conserved vector arising due to the translational symmetry of the system. Such a continuous dependence on the parameter $r_0$ leads to a band structure of the exciton spectrum.}
    \label{pic2_exciton_integrals}
\end{figure}

Let us discuss here the quantitative features of the energy spectrum of the third panel in Fig.~\ref{fig:pic1_gaAs_30}. First, we observe the regions of a discrete spectrum and continuous bands. The discrete levels correspond to the trion states, while the lower boundary of each continuous region (blue, violet, orange, pink) relates to the onset of the exciton continuum. For instance, in the case $n_{h}=0$ and $n_{e}=1$, there are no discrete levels. To find the lower bound of each continuous region, we consider the problem involving non-interacting exciton and electron (see Appendix~\ref{sec:X_e}). Let us illustrate this approach for $(n_e, n_h) = (0,0)$ and $(0,1)$. In these simplest cases all calculations  are reduced to the analytical formula instead of diagonalization of matrix. The lower border $E^{\textup{X}}_{00}$ for the region $(0,0)$ can be found by means of the following relation:
\begin{align}
E^{\textup{X}}_{00} &= E^{(\textup{X},0)}_{00}+V_{\vb{K}\rightarrow \vb{K}_{\textup{min}},00; 00} \nonumber \\
{}& = \frac{1}{2} \hbar \omega_0 + \left.I^{(\textup{X},eh,\textup{C})}_{00}\right\rvert_{r_0\rightarrow r_{\textup{min}}},\label{eqn:ex_e_ex_min_00}
\end{align}
where $E^{(\textup{X},0)}_{00}$ is the corresponding free LL [Eq.~\eqref{eq:eh_energy}], $V_{\vb{K},00; 00}$ is the Coulomb matrix element [Eq.~\eqref{eqn:app_vknm}], $\omega_0 = \omega_e + \omega_h$, and $I^{(\textup{X},eh,\textup{C})}_{00}$ is defined in Eq.~\eqref{eqn:app_basic_interaction_ex_Coulomb}. As was mentioned above, since an exciton is a neutral system, the translational symmetry gives rise to a conservation law for vector $\vb{K}$. Then we introduce $\vb{r}_0 = -c(\vb{B} \times \vb{K})/(eB^2)$ and vary the magnitude of $\vb{r}_0$ because the energy is independent of its direction (see Appendix~\ref{sec:app_exciton}). In Eq.~\eqref{eqn:ex_e_ex_min_00}, the energy as a function of $r_0$ is minimal (see Fig.~\ref{pic2_exciton_integrals}). For $(n_e, n_h) = (0,0)$, one finds $r_\textup{min} = 0$. In the case $(1,0)$, we have the following relation:
\begin{align}
E^{\textup{X}}_{01} &=E^{(\textup{X},0)}_{01}+V_{\vb{K}\rightarrow \vb{K}_{\textup{min}},01; 01} \nonumber \\ 
{}& = \frac{1}{2} \big (\hbar \omega_e + 3 \hbar \omega_h \big ) + \left. 2I^{(\textup{X},eh,\textup{C})}_{11}\right\rvert_{r_0\rightarrow r_{\textup{min}}},\label{eqn:ex_e_ex_min_01}
\end{align}
where factor $2$ in the last term comes from the coefficient $B^{01}_{10}$ introduced in Eq.~\eqref{eqn:app_psi_nm_operators_explicit}. In Fig.~\ref{pic2_exciton_integrals} we display the functions $I^{(\textup{X},eh,\textup{C})}_{00}$, and  $2I^{(\textup{X},eh,\textup{C})}_{11}$ versus the dimensionless parameter $r_0/\lambda$ in units of $E_0 = \sqrt{\pi/2}(e^2/\varepsilon\lambda)$, which represents a characteristic energy scale of all of the matrix elements. Since the Coulomb contributions to the exciton energy can be continuously varied by changing the parameter $r_0$, the exciton spectrum contains continuous bands, which are clearly seen in the third panel of Fig.~\ref{fig:pic1_gaAs_30}. For GaAs at $B=30~\textup{T}$, we obtain $E^{\textup{X}}_{00}=28.665~\textup{meV}$ and $E^{\textup{X}}_{01}=48.207~\textup{meV}$.

\section{Rules for filling the trion Hamiltonian}\label{app:sec_fill_ham}

\begin{figure*}
    \centering
\includegraphics[width = 0.8\linewidth]{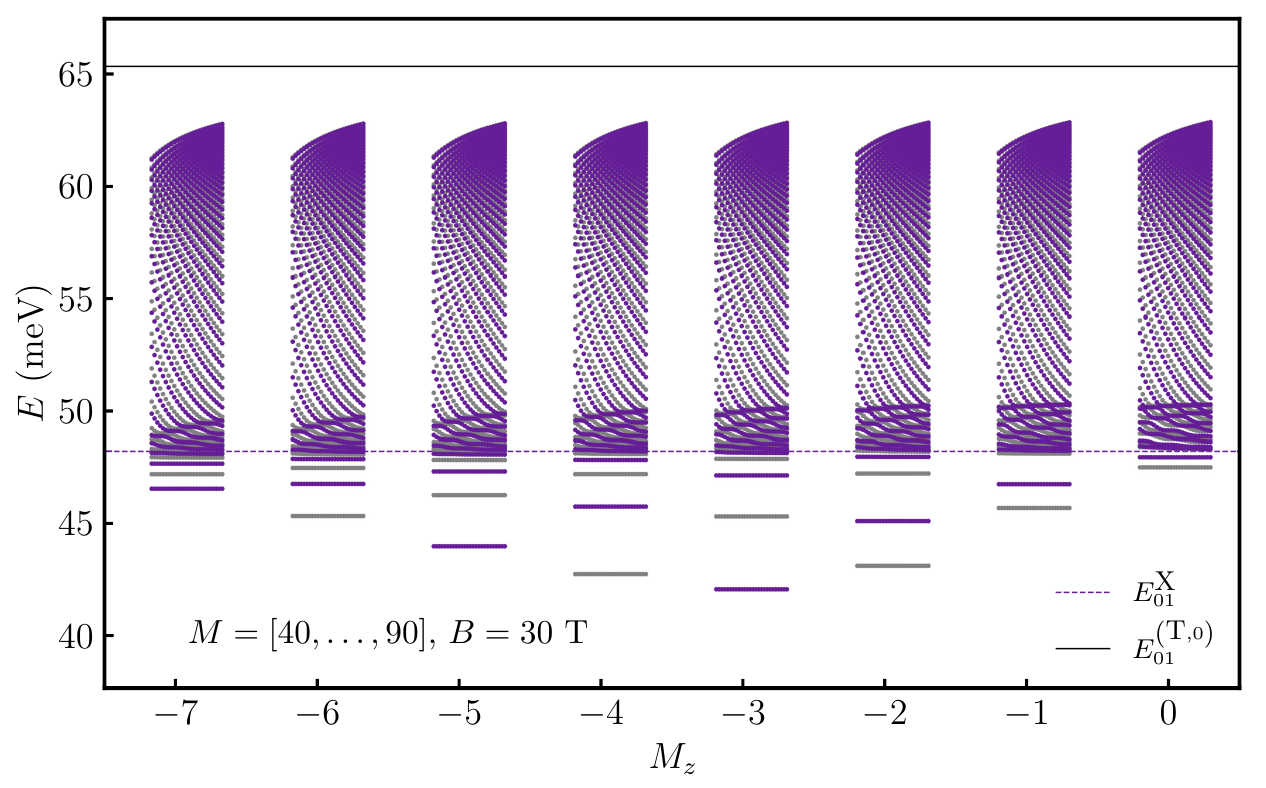}
    \caption{Detailed spectrum which corresponds to the  second~(purple) part of the third vertical panel in Fig.~\ref{fig:pic1_gaAs_30} for GaAs and $B=30~\textup{T}$. The points are grouped in accordance with specific values of $Mz$ (changes along the $x$ axis). The gray points are associated with triplets ($S_e=1$), while purple points correspond to singlets $S_e=0$. Recall that the indices associated with the free LLs are fixed and have the following values: $n_e=0$, $n_h=1$. For each integer $M_z$ we have $50$ points for each energy level. }
    \label{fig:pic7_GaAs_MZ_det_Dz}
\end{figure*}
\begin{figure*}
    \centering
\includegraphics[width = 0.8\linewidth]{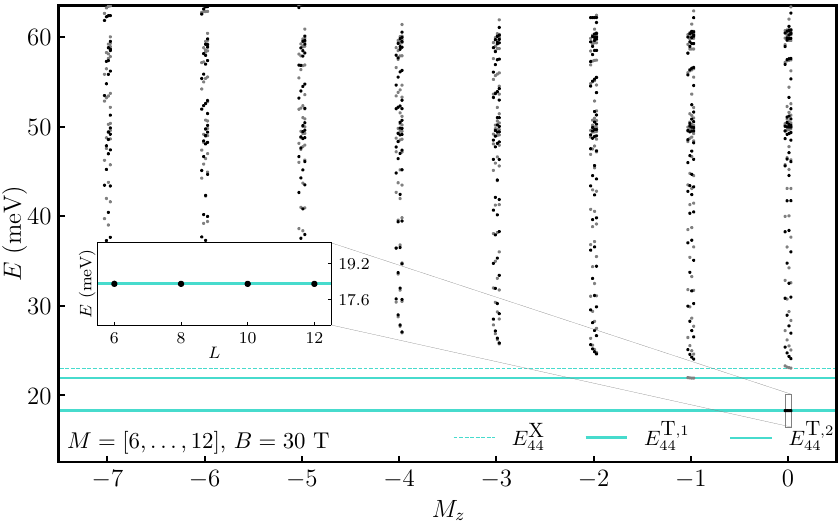}
    \caption{Detailed spectrum which corresponds to the lower part of the fourth vertical panel in Fig.~\ref{fig:pic1_gaAs_30} for GaAs and $B=30~\textup{T}$. The points are grouped in accordance with specific values of $Mz$ (changes along the $x$ axis). The gray points are associated with triplets ($S_e=1$), while black points correspond to singlets $S_e=0$. The indices associated with the free LLs change as $n_e=0\dots 4$, $n_h=0\dots 4$. For each integer $M_z$ we have four points for each energy level.}
    \label{fig:pic6_GaAs_MZ_det_0_3}
\end{figure*}

In order to obtain the spectrum of a trion, it is necessary to construct the Hamiltonian correctly. Let us look at the filling of the Hamiltonian using the example of the Fig.~\ref{fig:pic1_gaAs_30}. First, we analyze the third panel, where the matrix~\eqref{eq:HT} is  diagonalized neglecting the matrix elements between different LLs. Here, we employ the method proposed in Refs.~\cite{PhysRevLett.84.4429, DZYUBENKO2000683,PhysRevB.65.035318}. In practice, the matrix is filled in such a way that we do not get negative values of the indices, and also keep in mind that the parity and the projection of the total orbital angular momentum are preserved. Thus, if the arbitrary matrix element is given by $H^{(\textup{T})}_{n_1 n_2 n_h m l; n_1' n_2' n_h' m' l'}$, then the indices should satisfy the following conditions: $n_2=n_e-n_1$, $n_2'=n_e'-n_1'$, $m=\tilde{m} + (1 + (-1)^{n_1 + S_e})/2$, $m'=\tilde{m}' + (1 + (-1)^{n_1' + S_e})/2$, $l=\tilde{m} + (1 + (-1)^{n_1 + S_e})/2 + M_z - n_e + n_h$, $l'=\tilde{m}' + (1 + (-1)^{n_1' + S_e})/2 + M_z - n_e' + n_h'$, where $n_e$ and $n_e'$ change in the range $[n_{e,\textup{min}}, n_{e,\textup{max}}]$ with step 1, $n_1$ and $n_1'$ change in the range $[0, n_{e,\textup{max}}]$, $n_h$ and $n_h'$ change in the range $[n_{h,\textup{min}}, n_{h,\textup{max}}]$ with step 1, $\tilde{m}$ and $\tilde{m}'$ change in the range $[0, M]$ with step 2. Moreover, the following additional requirements should be verified:
$n_e - n_1\geq 0$,  $n_e' - n_1' \geq 0$,  $\tilde{m} + (1 + (-1)^{n_1 + S_e})/2 + M_z - n_e + n_h \geq 0$, $\tilde{m}' + (1 + (-1)^{n_1' + S_e})/2 + M_z - n_e' + n_h' \geq 0$.  The parameter $M$ is responsible for the convergence of the results affecting the size of the model Hilbert space. 

Thus, we should choose the following parameters: the value of the projection of the angular momentum $M_z$, the spin quantum number $S_e=0$ or $S_e=1$, $M$ (for the calculations at third panel in Fig.~\ref{fig:pic1_gaAs_30}  it is up to $90$), and the range of the parameters associated with the free LLs ($n_{h,\textup{min}}$, $n_{h,\textup{max}}$, $n_{e,\textup{min}}$, $n_{e,\textup{max}}$). For instance, in order to obtain the blue region of the spectra, we fix $n_{h,\textup{min}}=n_{h,\textup{max}}=n_{e,\textup{min}}=n_{e,\textup{max}}=0$, $M=90$ and collect the eigenvalues for different $M_z \in \{-7,\dots,0\}$ and $S_e=0$, $1$. The same steps we do for other colored parts presented in the third vertical panel in Fig.~\ref{fig:pic1_gaAs_30}, the difference is only in the choice of $n_{h,\textup{min}}$, $n_{h,\textup{max}}$, $n_{e,\textup{min}}$, $n_{e,\textup{max}}$. 

If the Coulomb mixing is taken into account, the corresponding spectrum is presented in the fourth vertical panel in Fig.~\ref{fig:pic1_gaAs_30}. Here, we chose $n_{h,\textup{min}}=n_{e,\textup{min}}=0$, and $n_{h,\textup{max}}=n_{e,\textup{max}}=4$, while $M$ is varied from $0$ to $12$. The smaller values of $M$ are employed due to the larger dimension of the model Hilbert space and corresponding computational complexity. Note that despite such a reduction in the dimension, this turns out to be sufficient for the convergence of the lower discrete levels.

To present a more detailed spectrum structure, we provided two examples of spectra where different parts corresponding to specific values of $M_z$ and $S_e$ are distinguished. One of them (Fig.~\ref{fig:pic7_GaAs_MZ_det_Dz}) relates to the purple part of the third vertical panel in Fig.~\ref{fig:pic1_gaAs_30}, while the second example (Fig.~\ref{fig:pic6_GaAs_MZ_det_0_3}) corresponds to the lower part of the fourth vertical panel in the same figure, where we take into account the Coulomb-induced mixing of the different LLs. The corresponding details can be found in Appendix~\ref{app:sec_spectrum_details}.

\section{Detailing of spectrum structure}\label{app:sec_spectrum_details}

Here we would like to illuminate the technicalities of our spectrum calculations. Of course, in practical trion computations, the regions that we associate with the exciton continuum are not continuous since the basis set employed is always finite, if huge. However, in order to correctly determine the continuous zones of the spectrum and establish the true discrete levels that we associate with the trion states, it is sufficient to (i) follow the behavior of the levels as the basis increases with increasing parameter $M$, (ii) carry out independent calculations of the lower exciton level in order to verify the lower border of the exciton continuum. Let us consider the first example in Fig.~\ref{fig:pic7_GaAs_MZ_det_Dz}. In order to construct it, we resolved the details of the purple sector of the third vertical panel in Fig.~\ref{fig:pic1_gaAs_30} with respect to the values of the projection of the total angular momentum $M_z$. The points are not merged into a single area as was done previously (in Fig.~\ref{fig:pic1_gaAs_30} we assigned a horizontal line of the same width to each point).
\begin{figure}[t!]
    \centering
    \includegraphics[width = 1.0\linewidth]{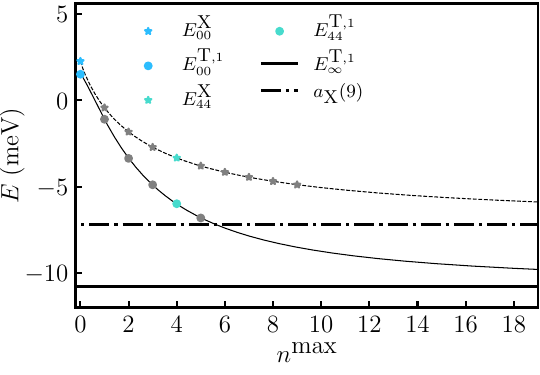}
    \caption{Dependence of the lowest trion level (gray circles) and lowest exciton level (gray stars) on the maximal value of the LLs taken into account when constructing the model Hilbert space in the case of GaAs at $B=10~\textup{T}$. Here we assume $n_e^{\textup{max}}=n_h^{\textup{max}}=n^{\textup{max}}$.}
    \label{fig:pic5_extr_GaAs_10}
\end{figure}
In Fig.~\ref{fig:pic7_GaAs_MZ_det_Dz}, we indicate the corresponding values of the projection $M_z$ below each group of the points at the horizontal axis. Blurring is due to the fact that when moving along the $x$ axis in the vicinity of each value of $M_z$, we change the value of the parameter $M$ demonstrating (i) the emergence of new levels, (ii) how the existing levels change their positions. Moreover, levels that have different spin number $S_e$ are marked by different colors (gray and purple). The spectrum shows two qualitatively different types of behavior. Above a certain level, with increasing parameter $M$, we observe a significant change in levels and the appearance of more and more new ones. However, in the region below the mentioned border, the levels practically do not change from the very beginning and stand apart from each other. Furthermore, new levels do not appear there. As one might guess, the upper part is associated with the exciton continuum, while the lower part yields the trion levels. Let us also note that the upper region tends to fill the entire part of the spectrum up to the corresponding free LL, which is also marked by the upper horizontal line.  As was already mentioned, to confirm this qualitative separation, we calculated the lower exciton level within the analysis of noninteracting exciton and electron (see Appendix~\ref{sec:X_e}). In Fig.~\ref{fig:pic7_GaAs_MZ_det_Dz} this level is labeled by $E^{\textup{X}}_{01}$. In Fig.~\ref{fig:pic6_GaAs_MZ_det_0_3} we show another example of detailing the projection of the total angular momentum $M_z$ and $S_e$ in the case of a more complicated consideration when mixing of various free LLs is taken into account. In order to obtain a clear structure of the discrete levels (trion ones), we limit ourselves only by the lower part of the spectrum presented in Fig.~\ref{fig:pic1_gaAs_30} within the fourth vertical panel. 
\begin{figure}[b]
    \centering
    \includegraphics[width = 1.0\linewidth]{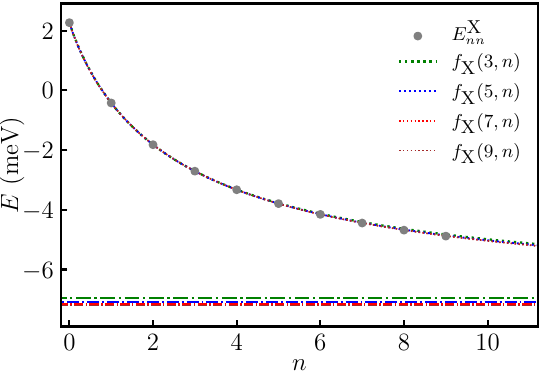}
    \caption{Behavior of the lowest exciton level with growth of Hilbert space (gray dots) for GaAs at $B=10~\textup{T}$. By the increasing we mean the number of LLs $n$ taken into account. These points are fitted using different extrapolations, built on the basis of the data of different length. The horizontal (dashed-dotted) lines indicate the values of corresponding $a_{\textup{X}}(n_{\textup{max}})$, for $n_{\textup{max}}=3,5,7,9$.}
    \label{fig:pic9_app_extr_ex}
\end{figure}
It is evident that with the inclusion of higher values of the free LLs $n_e$ and $n_h$, the dimension of the model Hilbert space grows significantly. For this reason, we did not achieve such values of $M$ as in the previous case. However, as it turned out, this was not necessary. Moreover, it is easier to identify the discrete levels since they demonstrate stability with respect to increasing $M$ from the very beginning. An example of such stability is demonstrated for the two lower trion levels. Moreover, for the lowest one, we display an inset in Fig.~\ref{fig:pic6_GaAs_MZ_det_0_3} which clearly shows the converged value of the energy level.

\section{Convergence of energy levels and extrapolation procedure}\label{app:sec_extrapol}

Our predictions should converge with respect to the number of the LLs incorporated within the model Hilbert space. It turns out that even within the setup that takes into account all levels up to the fifth ones ($n_e=0,\dots,5$ and $n_h=0,\dots,5$), we cannot conclude that the convergence has been achieved. In order to overcome this problem, we utilize an extrapolation procedure described below. In Fig.~\ref{fig:pic5_extr_GaAs_10}, we demonstrate the energy of one of the lowest trion levels and the lowest exciton level as functions of the number of LLs included within the model Hilbert space in the case of GaAs at $B=10~\textup{T}$. We depict first the energies calculated in the zero approximation ($n^{\textup{max}}_e=n^{\textup{max}}_h=0$, blue dots), as was done in Ref.~\cite{PhysRevLett.84.4429}, and then we present the results computed via our general technique: for the trion problem we reached $n^{\textup{max}}_e=n^{\textup{max}}_h=5$, while for the exciton problem we took into acoount the states up to $n^{\textup{max}}_e=n^{\textup{max}}_h=9$. One observes that the convergence is very slow and, based on the existing number of points, it is difficult to precisely determine the values of the true trion and exciton levels. However, using these points, we performed an extrapolation procedure allowing us to obtain more accurate predictions. In Fig.~\ref{fig:pic5_extr_GaAs_10}, they are depicted by the horizontal lines. We note that although extrapolating the data can never yield absolutely accurate values, it is not possible to extract the energy levels without extrapolation since the convergence is unlikely to be achieved by a brute-force approach, i.e, by simply increasing the dimension of the model Hilbert space via including additional LLs, due to the slow convergence rate and computational limitations.

\begin{figure*}[t!]
\centering
\includegraphics[width = 0.8\linewidth]{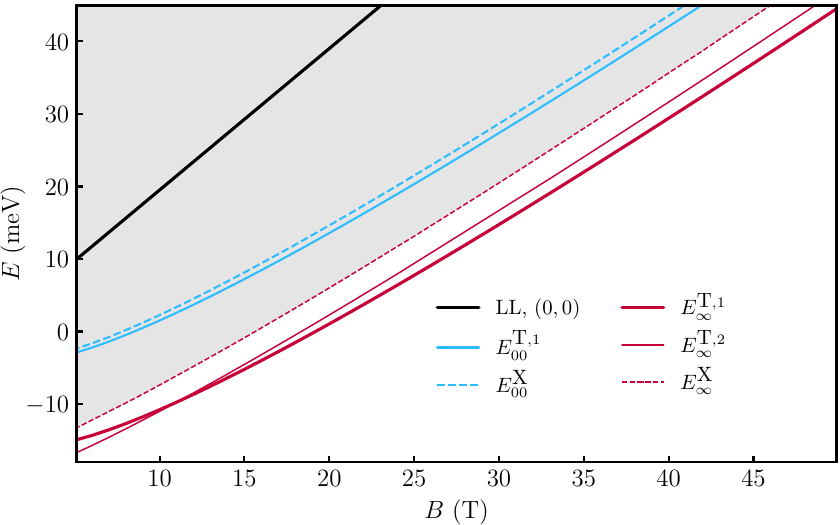}
\caption{The dependence of the energy spectrum of the system as function of the magnetic field B. Solid red lines correspond to the two discrete trion states, that we get in our calculation the dashed red line corresponds to the lower boundary of the energy continuum solid blue line corresponds to the trion bound state reported in Ref.~ \cite{DZYUBENKO2000683}, and dashed blue line to corresponding energetic continuum boundary. Solid black line shows the position of the lowest Landau level corresponding to non-interacting particles. 
The increase of magnetic field does not lead to the merging of red and blue lines, which means, that the role of the Coulomb mixing between different Landau levels does not decrease with magnetic field.}
\label{fig:pic6_GaAs_levels_vs_B}
\end{figure*}

Thus, to overcome this problem, we extrapolate the energy values in order to determine more accurate estimates. This procedure is obviously not unambiguous, so the predictions will always possess, in fact, some uncertainties. Here we will describe the approach we used for Fig.~\ref{fig:pic6_GaAs_levels_vs_B}. Below, we consider one point where $B=10~\textup{T}$. 
\begin{figure}[t]
    \centering
    \includegraphics[width = 1.0\linewidth]{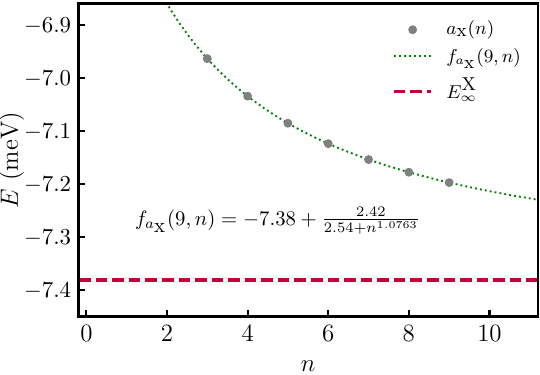}
    \caption{Gray dots show the behavior of the free constant $a_{\textup{X}}(n)$ entering extrapolation formula~\eqref{app:eqn_extrapol_formula_exciton} with the growth of Hilbert space (increasing the number of taken into account LLs). The dotted green line presents the highest extrapolation $f_{a_{\textup{X}}}(9,n)$ obtained on the basis of all gray dots, while the red dashed line is the asymptotic  value of the corresponding free constant which is interpreted by us as the final estimate for $E^{X}_{\infty}$. }
    \label{pic10_app_extr_ex_a_const}   
\end{figure}
\begin{figure}[t]
    \centering
    \includegraphics[width = 1.0\linewidth]{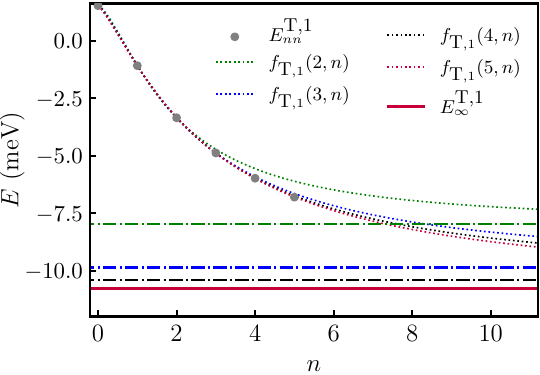}
    \caption{Gray dots demonstrate the behavior of one of the discrete trion levels with growth of Hilbert space (increasing of $n_{\textup{max}}$). These points are fitted using different extrapolations, built on the basis of the data of different length. The horizontal (dashed-dotted) lines indicate the values of corresponding $a_{\textup{T}}(n_{\textup{max}})$, for $n_{\textup{max}}=2,3,4$. The value of $a_{\textup{T}}(5)$ is identified separately by solid red line and considered by us as the final estimate of  $E^{\textup{T,}1}_{\infty}$ at fixed $B$.}
    \label{fig:pic11_app_extr_tr_1}
\end{figure}

The basis size is governed by the number $n_{\textup{max}}$, which is a maximal value of the LL quantum numbers $n_e$ and $n_h$. Let us first analyze the behavior of the lowest exciton level. The change in its position as a function of $n_{\textup{max}}$ is shown in Fig.~\ref{fig:pic9_app_extr_ex} (see gray dots). An important issue here is the choice of the form of the approximation function. Our analysis of the data obtained suggests that the natural approximation reads
\begin{align}\label{app:eqn_extrapol_formula_exciton}
    \!\!f_{\textup{X}}(n_{\textup{max}},n)\!=\!a_{\textup{X}}(n_\textup{max})+\dfrac{b_{\textup{X}}(n_\textup{max})}{n^{k_{\textup{X}}(n_\textup{max})}+c_{\textup{X}}(n_\textup{max})}.
\end{align}
It is clear that the fitting constants should alter with the appearance of new data points regarding $n_{\textup{max}}$. If these constants converge relatively well, we could conclude that the approximation formula is chosen correctly. Here we are only interested in the parameter $a_{\textup{X}}$ and its asymptotic value $a_{\textup{X}}(n\rightarrow\infty)$ which gives us the final estimate for the lowest exciton level.

In Fig.~\ref{pic10_app_extr_ex_a_const}, we present values of $a_{\textup{X}}(n_{\textup{max}})$ involved in the different extrapolation formulas $f_{\textup{X}}(n_{\textup{max}},n)$. Since the dimension of the model Hilbert space increases, the value of the constant itself also changes but the shifts are observed in a convergent manner. We construct a similar extrapolation for $a_{\textup{X}}(n_{\textup{max}})$ which we plot in the figure. The corresponding extrapolation function is labeled by $f_{a_{\textup{X}}}$. In the figure we also depict the red line which gives the asymptotic value $a_{\textup{X}}(n\rightarrow\infty)$ and is considered as a final estimate for $E^{\textup{X}}_{\infty}$. The same steps were applied for extracting $E^{\textup{X}}_{\infty}$ for other values of the magnetic field $B$ presented in Fig.~\ref{fig:pic6_GaAs_levels_vs_B}.

Finally, let us analyze the behavior of one of the lowest trion level. The dependence of its position as a function of $n_{\textup{max}}$ is demonstrated in Fig.~\ref{fig:pic11_app_extr_tr_1}. In Table~\ref{tab:1}, we also provide the numbers we will extrapolate.
\begin{table}[b]
 \centering
    \caption{Values of one of the lowest trion levels calculated at $B=10~\textup{T}$ for GaAs. The numbers correspond to different values of the LLs quantum number $n$. The corresponding values of the conserved quantities $S_e$ and $M_z$ are also shown.} %
    \label{tab:1}
     \setlength{\tabcolsep}{6.24pt}
    \begin{tabular}{rrrr|rrrr}
      \hline
      \hline
      $n$ & $E^{\textup{T,}1}_{nn}\textup{(meV)}$& $M_z$ & $S_e$  &     $n$ & $E^{\textup{T,}1}_{nn}\textup{(meV)}$& $M_z$ & $S_e$ \\ 
      \hline
        $0$  &    $1.5172$  & $-1$  & $1$     &        $3$  &    $-4.8842$  & $0$  & $0$     \\
        $1$  &    $-1.0890$  & $0$  & $0$    &        $4$  &    $-5.9807$  & $0$  & $0$     \\ 
        $2$  &    $-3.3553$  & $0$  & $0$    &        $5$  &    $-6.8054$  & $0$  & $0$     \\
    \hline
    \hline
    \end{tabular}
\end{table}
The form of the approximation function is chosen the same way as in the exciton case:
\begin{align}
\!\!    f_{\textup{T},1}(n_{\textup{max}},n)\!=\!a_\textup{T}(n_\textup{max})+\dfrac{b_\textup{T}(n_\textup{max})}{n^{k_\textup{T}(n_\textup{max})}+c_\textup{T}(n_\textup{max})}.
\end{align}
As in the case of the lowest  exciton level, it is necessary to monitor the convergence of the extrapolation constants when $n_{\textup{max}}$ increases. If they converge relatively well, we can conclude that the approximation formula is chosen correctly. We are only interested in the parameter $a_{\textup{T}}$. For the approximations that are available to us, we extract the following values: $a_\textup{T}(2)=-7.9902~\textup{meV}$,  $a_\textup{T}(3)= -9.8591~\textup{meV}$, $a_\textup{T}(4)=-10.400~\textup{meV}$, $a_\textup{T}(5)=-10.783~\textup{meV}$. Unfortunately, in contrast to the exciton problem, such a number of points is not enough to construct an extrapolation for the behavior of the constant itself; here we decided to limit ourselves to only its upper estimate and consider $a_\textup{T}(5)=-10.783~\textup{meV}$ as the final estimate for the lowest trion level $E^{\textup{T},1}_{\infty}$ for $B=10~\textup{T}$. The same procedure is repeated for all points in Fig.~\ref{fig:pic6_GaAs_levels_vs_B}

Finally, let us discuss the intersection between the two trion levels revealed in Fig.~\ref{fig:pic6_GaAs_levels_vs_B}. First, we note that they may indeed intersect since they have different values of the spin quantum number $S_e$. Second, we underline that the exact treatment of the three-particle problem is very complicated from the computational viewpoint, so the quantitative predictions are, in fact, not sufficiently accurate to properly resolve the relative position of the two levels. Although one can evidently observe both of them, it might be difficult to identify the corresponding energy difference for a given value of the magnetic field strength and, therefore, to accurately localize the intersection point. Providing a high-resolution picture represents a formidable task and an important challenge for future studies.

\section{\textit{Mathematica} subroutine}\label{app:math_sr}

As Supplemental Material, we attach \textit{Mathematica} files with the help of which one can calculate any matrix elements and the spectra presented in the paper.  The archive contains two files: \texttt{func\_spec.m} which includes basic functions and a \textit{Mathematica} notebook \texttt{Trion\_exciton\_spectrum\_in\_megnetic\_field.nb} which contains examples of calculations. To speed up the computations, we calculated an extensive database of the matrix elements for exciton and trion problems, which are located in directories \texttt{data\_ex} and \texttt{data\_tr}.

\bibliography{lit.bib}
\end{document}